\documentclass[pra,floatfix,twocolumn,amsfonts,amsmath,amssymb,nofootinbib,reprint,superscriptaddress]{revtex4-2}
\pdfoutput=1
\usepackage[utf8]{inputenc}
\usepackage[T1]{fontenc}
\usepackage{accents}
\usepackage[colorlinks]{hyperref}
\usepackage[normalem]{ulem}

\usepackage{amsthm}
\usepackage{mathtools}
\usepackage{bbm}
\usepackage{bm}
\usepackage{graphicx}

\newtheorem{theorem}{Theorem}

\newtheorem{definition}[theorem]{Definition}

\def\bra#1{\mathinner{\langle{#1}|}}
\def\ket#1{\mathinner{|{#1}\rangle}}
\def\braket#1#2{\mathinner{\langle{#1|#2}\rangle}}
\def\ketbra#1#2{\mathinner{|{#1}\rangle\!\langle{#2}|}}

\def\dbra#1{\mathinner{\langle\!\langle{#1}|}}
\def\dket#1{\mathinner{|{#1}\rangle\!\rangle}}
\def\dketbra#1#2{\mathinner{|{#1}\rangle\!\rangle\!\langle\!\langle{#2}|}}

\newcommand{\id}{\mathbbm{1}}

\DeclareMathOperator{\Tr}{Tr}
\DeclareMathOperator{\range}{range}

\renewcommand{\L}{\mathcal{L}}
\renewcommand{\P}{\mathcal{P}}
\renewcommand{\S}{\mathcal{S}}
\newcommand{\M}{\mathcal{M}}
\newcommand{\HS}{\mathcal{H}}

\newcommand{\K}{\mathcal{K}}

\newcommand{\E}{\mathcal{E}}

\newcommand{\W}{\mathcal{W}}

\newcommand{\I}{\mathcal{I}}

\begin{document}

\title{Semi-Device-Independent Certification of Causal Nonseparability\\
with Trusted Quantum Inputs}

\author{Hippolyte Dourdent}
\affiliation{Univ.\ Grenoble Alpes, CNRS, Grenoble INP\footnote{Institute of Engineering Univ. Grenoble Alpes}, Institut N\'eel, 38000 Grenoble, France}

\author{Alastair A.\ Abbott}
\affiliation{Univ.\ Grenoble Alpes, Inria, 38000 Grenoble, France}
\affiliation{D\'epartement de Physique Appliqu\'ee, Universit\'e de Gen\`eve, 1211 Gen\`eve, Switzerland}

\author{Nicolas Brunner}
\affiliation{D\'epartement de Physique Appliqu\'ee, Universit\'e de Gen\`eve, 1211 Gen\`eve, Switzerland}

\author{Ivan Šupić}
\affiliation{CNRS, LIP6, Sorbonne Université, 4 Place Jussieu, 75005 Paris, France}
\affiliation{D\'epartement de Physique Appliqu\'ee, Universit\'e de Gen\`eve, 1211 Gen\`eve, Switzerland}

\author{Cyril Branciard}
\affiliation{Univ.\ Grenoble Alpes, CNRS, Grenoble INP\footnote{Institute of Engineering Univ. Grenoble Alpes}, Institut N\'eel, 38000 Grenoble, France}

\date{August 26, 2022}

\begin{abstract}
While the standard formulation of quantum theory assumes a fixed background causal structure, one can relax this assumption within the so-called process matrix framework. Remarkably, some processes, termed causally nonseparable, are incompatible with a definite causal order. We explore a form of certification of causal nonseparability in a semi-device-independent scenario where the involved parties receive trusted quantum inputs, but whose operations are otherwise uncharacterised. Defining the notion of causally nonseparable distributed measurements, we show that certain causally nonseparable processes which cannot violate any causal inequality, including the canonical example of the quantum switch, can generate noncausal correlations in such a scenario. Moreover, by imposing some further natural structure to the untrusted operations, we show that all bipartite causally nonseparable process matrices can be certified with trusted quantum inputs.
\end{abstract}

\maketitle

When reasoning about quantum and classical processes alike, we usually assume a fixed causal structure.
Remarkably, this turns out to be an unnecessarily restrictive assumption: there are valid processes with indefinite causal order.
Such processes can be formalised within the process matrix framework, where quantum theory is taken to hold locally but no global causal structure is assumed~\cite{oreshkov12}.
The existence of processes incompatible with a definite causal order, termed ``causally nonseparable'', bears a foundational significance, but moreover can be the basis for advantages in a number of different tasks~\cite{chiribella12,araujo14,guerin16}.  

Some causally nonseparable process matrices can generate so-called noncausal correlations, allowing their causal nonseparability to be certified in a \textit{device-independent} (DI) way by violating ``causal inequalities''~\cite{oreshkov12,branciard16}. 
However, not all causally nonseparable process matrices are noncausal in this strong sense~\cite{araujo15,oreshkov16,feix16}.
Indeed, it remains unclear if any physically realisable process can violate a causal inequality, and causal models have recently been formulated for a large class of quantum-realisable processes~\cite{wechs21,purves21}. This notably includes the canonical ``quantum switch''~\cite{chiribella13}, the resource behind most known advantages arising from causal indefiniteness.
At the same time, causally nonseparable process matrices can always be certified by ``causal witnesses''~\cite{araujo15,branciard16a}.
This approach, however, has the drawback of being \textit{device-dependent} (DD), as it requires one to perfectly trust the operations performed by the involved parties. 

Given the obstacles towards employing a DI approach to certify particularly relevant processes, there is particular urgency in exploring intermediate, semi-DI (SDI) approaches. 
One possible approach recently considered is to trust only some of the parties' operations~\cite{bavaresco19}.
Here we explore a different SDI regime, significantly weakening the requirements of trust on all parties while simultaneously obtaining a widely applicable certification.
Inspired by recent developments in quantum nonlocality~\cite{buscemi12,branciard13}, we consider a causal game scenario where the parties receive inputs in the form of trusted quantum systems (instead of classical ones), but are otherwise untrusted or uncharacterised. 
We show that certain causally nonseparable processes which cannot violate any causal inequality, including the quantum switch~\cite{chiribella13,araujo15,oreshkov16}, can nevertheless display some new form of noncausality in a \emph{semi-DI with quantum inputs} (SDI-QI) scenario. 
We then consider a more constrained version of this scenario in which the uncharacterised operations have a specific, but rather natural structure, and we show that \emph{all} bipartite causally nonseparable process matrices can be certified in this ``measurement device and channel independent'' (MDCI) scenario.

\textit{Causal (non)separability in the process matrix framework.---}%
We focus initially on the bipartite scenario, before returning, toward the end of this Letter, to the more practically pertinent scenario in which the quantum switch is formulated. Two parties, Alice and Bob, control separate labs with input and output Hilbert spaces $\HS^{A_I}$ and $\HS^{A_O}$ for Alice, and $\HS^{B_I}$ and $\HS^{B_O}$ for Bob.
They may also receive some ancillary quantum states in Hilbert spaces $\HS^{\tilde{A}}$, $\HS^{\tilde{B}}$, $\rho^{\tilde{A}\tilde{B}} \in \L(\HS^{\tilde{A}\tilde{B}})$.
(Here and throughout, we denote the space of linear operators on $\HS^X$ as $\L(\HS^X)$ and write concisely $\HS^{XY} = \HS^X\otimes\HS^Y$, $\HS^A = \HS^{A_IA_O}$, etc.; 
superscripts indicate on what spaces operators act.)
They perform quantum operations described as quantum instruments~\cite{davies70}, i.e., sets of completely positive (CP) maps $\M_a: \L(\HS^{\tilde{A}A_I})\to\L(\HS^{A_O})$ and $\M_b: \L(\HS^{\tilde{B}B_I})\to\L(\HS^{B_O})$, whose indices $a,b$ refer to some (classical) outcomes for Alice and Bob, and whose sums $\sum_a \M_a$ and $\sum_b \M_b$ are trace-preserving (TP). 

Using the Choi isomorphism~\cite{choi75} (see Appendix~\ref{app:Choi_link_prod}), the CP maps $\M_a$, $\M_b$ can be represented as positive semidefinite (PSD) matrices $M_a^{\tilde{A}A}$ and $M_b^{\tilde{B}B}$.
Within the process matrix framework, the correlations established by Alice and Bob are then given by the probabilities
\begin{align}
P(a,b) & = \Tr \left[ \left(M_a^{\tilde{A}A}\otimes M_b^{\tilde{B}B}\right)^T \left(\rho^{\tilde{A}\tilde{B}}\otimes W^{AB}\right) \right],
\label{eq:gen_Born}
\end{align}
where $W^{AB} \in \L(\HS^{AB})$ is the so-called ``process matrix''. 
To ensure that Eq.~\eqref{eq:gen_Born} always defines valid probabilities, $W^{AB}$ must be PSD and belong to a nontrivial subspace of $\L(\HS^{AB})$~\cite{oreshkov12} (see Appendix~\ref{app:validity_cstrs}).

The process matrix formalism makes no \emph{a priori} assumption of a global causal structure relating Alice and Bob.
In fact, the assumption of such a structure imposes further constraints, due to the inability for a party to ``signal'' to the causal past.
Process matrices compatible, for example, with Alice acting before Bob (denoted $A\prec B$) are of the form $W^{A\prec B} = W^{A\prec B_I} \otimes \id^{B_O}$, and similarly $W^{B\prec A} = W^{B\prec A_I} \otimes \id^{A_O}$ for Bob before Alice ($B\prec A$), with $W^{A\prec B_I}$ and $W^{B\prec A_I}$ being themselves valid process matrices~\cite{oreshkov12}.
Process matrices that can be written as a convex mixture of matrices compatible with $A\prec B$ and $B\prec A$, i.e., of the form
\begin{align}
W^{AB} & = q\,W^{A\prec B_I} \otimes \id^{B_O} + (1{-}q)\,W^{B\prec A_I} \otimes \id^{A_O}
\label{eq:csep2}
\end{align}
with $q\in[0,1]$ are said to be \emph{causally separable}. 
They can be interpreted as being compatible with a definite (although probabilistic) causal order. 
Remarkably, there exist \emph{causally nonseparable} process matrices that cannot be decomposed as in Eq.~\eqref{eq:csep2}, and are thus incompatible with any definite causal order~\cite{oreshkov12}.

As recalled above, causal nonseparability can always be certified in a DD manner using a causal witness~\cite{araujo15,branciard16a}, while some processes can be certified in a DI way through the violation of a causal inequality~\cite{oreshkov12}. 
Here we consider a relaxation of the DI scenario, where rather than viewing the parties as black boxes with classical inputs and outputs, we provide them with quantum inputs.
This intermediate SDI-QI scenario has previously been shown to be extremely useful for entanglement certification~\cite{buscemi12}, but its applicability to causal nonseparability, where parties implement instruments rather than just measurements, remains unstudied.

\textit{Process matrix scenario with quantum inputs.---}%
We thus consider a situation where Alice and Bob are provided with quantum input states $\rho_x^{\tilde{A}}$ and $\rho_y^{\tilde{B}}$, respectively, indexed by the labels $x$ and $y$.
They each perform some fixed instruments $(M_a^{\tilde{A}A})_a$ and $(M_b^{\tilde{B}B})_b$. 
We explicitly write the dependency on the quantum inputs in the correlations $P(a,b|\rho_x^{\tilde{A}},\rho_y^{\tilde{B}})$ obtained according to Eq.~\eqref{eq:gen_Born}, with $\rho^{\tilde{A}\tilde{B}} = \rho_x^{\tilde{A}}\otimes\rho_y^{\tilde{B}}$.

\begin{figure}[t]
	\begin{center}
	\includegraphics[width=1\columnwidth]{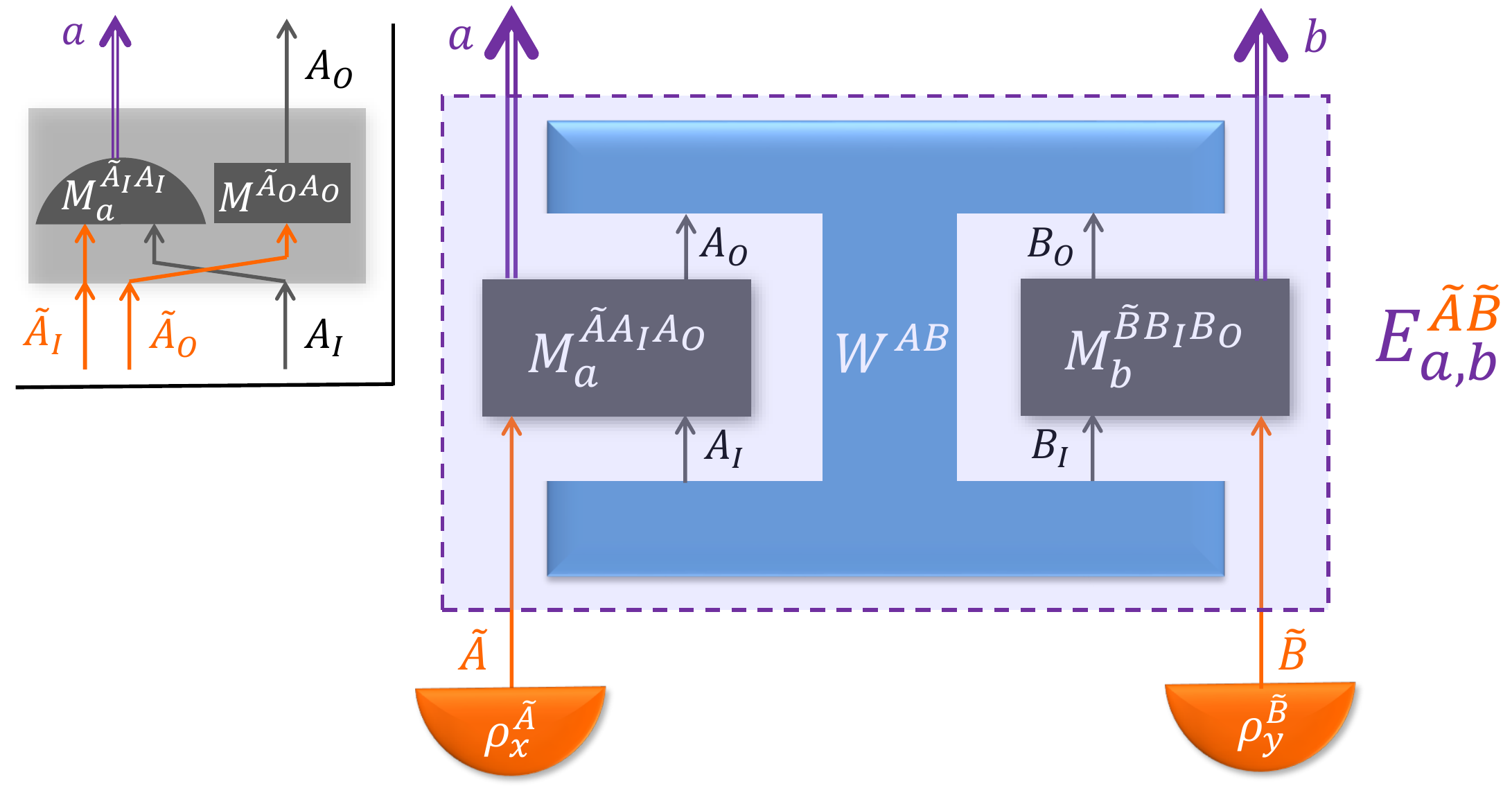}
	\end{center}
	\caption{SDI-QI scenario (main): A process matrix $W^{AB}$ connects two parties who receive quantum inputs $\rho_x^{\tilde{A}}$ and $\rho_y^{\tilde{B}}$, resp. They each perform a joint operation (($M_a^{\tilde{A}A_I A_O})_a$ and $(M_b^{\tilde{B}B_I B_O})_b$, resp.), and produce the classical outcomes $a$ and $b$. The purple box shows the D-POVM $(E_{a,b}^{\tilde{A}\tilde{B}})_{a,b}$ induced by these instruments and the process matrix. 
	Inset: In the MDCI scenario (see later), additional structure is assumed on the quantum instruments (shown here for Alice). The quantum input is a bipartite state in $\HS^{\tilde{A}_I\tilde{A}_O}$, a measurement is performed jointly on $\HS^{\tilde{A}_IA_I}$ and a channel sends $\HS^{\tilde{A}_O}$ to the process matrix through $\HS^{A_O}$.}
	\label{fig:DPOVM}
\end{figure}

It will be convenient in our calculations to use the so-called ``link product'' $*$~\cite{chiribella08,chiribella09}, defined for any matrices $M^{XY} \in \L(\HS^{XY})$, $N^{YZ} \in \L(\HS^{YZ})$ as $M^{XY}*N^{YZ} = \Tr_Y[(M^{XY}\otimes\id^Z)^{T_Y}(\id^X\otimes N^{YZ})] \in \L(\HS^{XZ})$ (where $T_Y$ is the partial transpose over $\HS^Y$; see also Appendix~\ref{app:Choi_link_prod}).
Noting that a full trace $\Tr[(M^Y)^T N^Y]$ and a tensor product $M^X \otimes N^Z$ can both be written as a link product, and that the link product is commutative and associative, Eq.~\eqref{eq:gen_Born} can be written as
\begin{align}
P(a,b|\rho_x^{\tilde{A}},\rho_y^{\tilde{B}}) & = \left(M_a^{\tilde{A}A}\otimes M_b^{\tilde{B}B}\right) * \left(\rho_x^{\tilde{A}}\otimes\rho_y^{\tilde{B}}\otimes W^{AB}\right) \notag \\
 & \hspace{-18mm} = E_{a,b}^{\tilde{A}\tilde{B}} * \left(\rho_x^{\tilde{A}}\otimes\rho_y^{\tilde{B}}\right) = \Tr \left[ \left(E_{a,b}^{\tilde{A}\tilde{B}}\right)^{\!T} \!\! \left(\rho_x^{\tilde{A}}\otimes\rho_y^{\tilde{B}}\right) \right]
\label{eq:Pab_rhoxrhoy}
\end{align}
with $E_{a,b}^{\tilde{A}\tilde{B}} = \left(M_a^{\tilde{A}A}\otimes M_b^{\tilde{B}B}\right)*W^{AB}$.
According to Eq.~\eqref{eq:Pab_rhoxrhoy}, the family $\mathbb{E}^{\tilde{A}\tilde{B}} \coloneqq (E_{a,b}^{\tilde{A}\tilde{B}})_{a,b}$ defines an effective, ``distributed'' measurement~\cite{supic17,hoban18} on the quantum inputs, which we term a ``distributed positive-operator-valued measure'' (D-POVM); see Fig.~\ref{fig:DPOVM}.

In the SDI-QI approach, the quantum inputs $\rho_x^{\tilde{A}}, \rho_y^{\tilde{B}}$ and their respective spaces are taken to be trusted.
However, we do not trust the instruments $(M_a^{\tilde{A}A})_a$ and $(M_b^{\tilde{B}B})_b$, and make no assumptions about the spaces $\HS^{A_I}, \HS^{A_O}, \HS^{B_I}$ and $\HS^{B_O}$.
Provided we can use a tomographically complete set of trusted quantum inputs, the D-POVM elements $E_{a,b}^{\tilde{A}\tilde{B}}$ can be explicitly reconstructed via Eq.~\eqref{eq:Pab_rhoxrhoy}.
The fundamental question we address here is this: if $W^{AB}$ is causally nonseparable, can one certify its causal nonseparability by just looking at the $E_{a,b}^{\tilde{A}\tilde{B}}$'s? 
To tackle this question, we ask conversely whether assuming that $W^{AB}$ is causally separable imposes any specific constraints on the $E_{a,b}^{\tilde{A}\tilde{B}}$'s.

\textit{Causally separable D-POVMs.---}%
Suppose that $W^{AB} = W^{A\prec B_I}\otimes \id^{B_O}$ is compatible with the order $A\prec B$. Then one can easily show (see Appendix~\ref{app:causallySepDPOVMs}) that
\begin{align}
\sum_b E_{a,b}^{\tilde{A}\tilde{B}} & = E_a^{\tilde{A}} \otimes \id^{\tilde{B}} \label{eq:nosig_E_AprecB}
\end{align}
with $E_a^{\tilde{A}} = M_a^{\tilde{A}A} * \Tr_{B_I} W^{A\prec B_I} \ge 0$ defining a (single-partite) POVM $(E_a^{\tilde{A}})_a$.
Eq.~\eqref{eq:nosig_E_AprecB} can be interpreted as a no-signalling condition from Bob to Alice~\cite{supic17}: indeed, it implies that Alice's marginal probability distribution does not depend on Bob's quantum input. 
A D-POVM satisfying $\sum_b E_{a,b}^{\tilde{A}\tilde{B}} = E_a^{\tilde{A}} \otimes \id^{\tilde{B}}$ for all $a$ is thus compatible with the causal order where Alice receives her quantum input and acts before Bob ($\tilde{A}\prec\tilde{B}$); we generically denote such a D-POVM $\mathbb{E}^{\tilde{A}\prec\tilde{B}} = (E_{a,b}^{\tilde{A}\prec\tilde{B}})_{a,b}$.
Similarly, for the order $B\prec A$, the resulting D-POVM must satisfy $\sum_a E_{a,b}^{\tilde{A}\tilde{B}} = \id^{\tilde{A}} \otimes E_b^{\tilde{B}}$ for all $b$; we generically denote such a D-POVM $\mathbb{E}^{\tilde{B}\prec\tilde{A}} = (E_{a,b}^{\tilde{B}\prec\tilde{A}})_{a,b}$. 

In analogy with the corresponding definition for process matrices (cf.\ Eq.~\eqref{eq:csep2}), we introduce the following:
\begin{definition}\label{def:csep_POVM_2}
A bipartite D-POVM $\mathbb{E}^{\tilde{A}\tilde{B}}$ that can be decomposed as a convex mixture of D-POVMs compatible with the causal orders $\tilde{A}\prec\tilde{B}$ and $\tilde{B}\prec\tilde{A}$, i.e., of the form
\begin{align}
\mathbb{E}^{\tilde{A}\tilde{B}} = q\, \mathbb{E}^{\tilde{A}\prec\tilde{B}} + (1{-}q)\, \mathbb{E}^{\tilde{B}\prec\tilde{A}} \label{eq:csep_POVM_2}
\end{align}
with $q\in[0,1]$ is said to be \emph{causally separable}.
\end{definition}

Clearly, it follows from the previous discussion that a causally separable process matrix can only generate causally separable D-POVMs. 
It turns out (see Appendix~\ref{app:any_csep_DPOVM}) that the converse also holds: any causally separable D-POVM can be realised by appropriate local operations on a causally separable process matrix.

\textit{SDI-QI certification of causal nonseparability.---}%
Let us note already that one can verify whether a given D-POVM is causally nonseparable with semidefinite programming (SDP).
Just as for process matrices~\cite{araujo15,branciard16a}, one can indeed construct ``witnesses of causal nonseparability for D-POVMs'' that certify any causally nonseparable D-POVM $\mathbb{E}^{\tilde{A}\tilde{B}}$ (see Appendix~\ref{app:cones_charact}).
Concretely, a witness provides a family $\mathbb{S}^{\tilde{A}\tilde{B}}=(S_{a,b}^{\tilde{A}\tilde{B}})_{a,b}$ of operators such that $\sum_{a,b}S_{a,b}^{\tilde{A}\tilde{B}} * E_{a,b}^{\tilde{A}\tilde{B}} < 0$ only if $\mathbb{E}^{\tilde{A}\tilde{B}}$ is causally nonseparable.
Taking $\{\rho_x^{\tilde{A}}\}_x$ and $\{\rho_y^{\tilde{B}}\}_y$ to be tomographically complete sets and writing $S_{a,b}^{\tilde{A}\tilde{B}}=\sum_{x,y}s_{a,b}^{(x,y)}\rho_x^{\tilde{A}}\otimes\rho_y^{\tilde{B}}$, one can thus reconstruct the witness from the correlations $P(a,b|\rho_x^{\tilde{A}},\rho_y^{\tilde{B}})$ and certify the causal nonseparability of $\mathbb{E}^{\tilde{A}\tilde{B}}$ by observing
\begin{equation}
	\sum_{a,b}S_{a,b}^{\tilde{A}\tilde{B}} * E_{a,b}^{\tilde{A}\tilde{B}} = 
	\sum_{a,b,x,y} s_{a,b}^{(x,y)} P(a,b|\rho_x^{\tilde{A}},\rho_y^{\tilde{B}}) < 0. \label{eq:reconstruct_witness}
\end{equation}

To certify the causal nonseparability of a process matrix in an SDI-QI manner, the key problem is thus to find some ancillary systems $\HS^{\tilde{A}}, \HS^{\tilde{B}}$ and some instruments $(M_a^{\tilde{A}A})_a$ and $(M_b^{\tilde{B}B})_b$ such that the D-POVM $\mathbb{E}^{\tilde{A}\tilde{B}}$ introduced in Eq.~\eqref{eq:Pab_rhoxrhoy} is causally nonseparable.

The simplest case is if a bipartite process matrix can generate noncausal correlations---i.e., if it is ``noncausal'', or even ``not extensibly causal''~\cite{oreshkov16}---then it is fairly easy to see that it can generate a causally nonseparable D-POVM.
Indeed, these processes can be certified in a fully DI manner through the violation of a causal inequality using classical, rather than quantum, inputs (cf.\ Appendix~\ref{app:noncausal_imply_nonsepDPOVM}).

Conceptually, it is more interesting to determine whether some ``causal'' process matrices can generate causally nonseparable D-POVMs.
One such bipartite process was formulated by Feix \emph{et al.}~\cite{feix16}.
We were again able to find simple instruments that directly generate a causally nonseparable D-POVM from this process (see Appendix~\ref{app:Feix}). 
In contrast, the alternative SDI approach of Ref.~\cite{bavaresco19} in which only some parties are trusted was unable to certify the causal nonseparability of this process. This highlights the potential power of our SDI-QI approach.

\textit{Certifying all bipartite causally nonseparable process matrices with trusted quantum inputs.---}%
The fact that the nonseparability of some specific causal processes can be nontrivially certified in an SDI-QI way leads one to wonder whether there is a systematic way to obtain a causally nonseparable D-POVM from any causally nonseparable process matrix.
Indeed, in the study of entanglement one can certify \emph{any} entangled state with trusted quantum inputs in a ``measurement-device-independent'' (MDI) manner, and a general recipe is known to construct MDI entanglement witnesses (MDIEWs)~\cite{buscemi12,branciard13}.
Currently this remains an open question with the general SDI-QI approach introduced above. 

Interestingly, the answer turns out to be positive, in the bipartite case, if 
one makes a further, physically motivated, assumption on the structure of the instruments used by Alice and Bob.
In particular, let us now consider a modified scenario, which we term ``measurement device and channel independent'' (MDCI) and where we assume that Alice and Bob's (trusted) ancillary Hilbert spaces have a bipartite structure of the form $\HS^{\tilde{A}} = \HS^{\tilde{A}_I\tilde{A}_O}$ and $\HS^{\tilde{B}} = \HS^{\tilde{B}_I\tilde{B}_O}$, and that their instruments have the following structure (here, e.g., for Alice; see also Fig.~\ref{fig:DPOVM} inset):
(i) Alice performs a joint quantum measurement (i.e., a POVM) on the subsystem of her quantum input in $\HS^{\tilde{A}_I}$ and the (untrusted) system in $\HS^{A_I}$ she receives from the process matrix; 
(ii) the part of the quantum input in $\HS^{\tilde{A}_O}$ is sent (independently from the joint measurement on $\HS^{\tilde{A}_IA_I}$) to the process matrix in the (untrusted) output space $\HS^{A_O}$ via a quantum channel (i.e., a CPTP map). 
The Choi maps of the instruments then factorise accordingly as
\begin{align}
M_a^{\tilde{A}A} & = M_a^{\tilde{A}_I A_I} \otimes M^{\tilde{A}_OA_O}, \quad
M_b^{\tilde{B}B} = M_b^{\tilde{B}_I B_I} \otimes M^{\tilde{B}_OB_O},  \label{eq:cstr_instr_v0}
\end{align}
with $\sum_a M_a^{\tilde{A}_I A_I} = \id^{\tilde{A}_I A_I}$ and $\Tr_{A_O} M^{\tilde{A}_OA_O} = \id^{\tilde{A}_O}$, and similarly for Bob. 
Importantly, in this MDCI scenario, we make no assumption about the POVMs and CPTP maps themselves, so they may be completely uncharacterised. 
We only assume the specified bipartite structure of the instruments, a natural assumption that can be physically justified if the quantum input is provided as two physically distinct systems (e.g., photons in two separate fibres) and distinct operations performed on these inputs. 

Using this additional structure, we prove in Appendix~\ref{app:MDCI_cert} that every element $E_{a,b}^{\tilde{A}\tilde{B}}$ of a D-POVM obtained from a causally separable process matrix $W^{AB}$ necessarily decomposes as
\begin{align}
E_{a,b}^{\tilde{A}\tilde{B}} & = q\,E_{a,b}^{\tilde{A}\prec\tilde{B}_I} \otimes \id^{\tilde{B}_O} + (1{-}q)\,E_{a,b}^{\tilde{B}\prec\tilde{A}_I} \otimes \id^{\tilde{A}_O} \label{eq:csep_POVM_00}
\end{align}
for some $E_{a,b}^{\tilde{A}\prec\tilde{B}_I}, E_{a,b}^{\tilde{B}\prec\tilde{A}_I} \ge 0$.
Remarkably, this structure is sufficient to certify the causal nonseparability of any causally nonseparable process matrix by looking at a single D-POVM element in a systematic way.
In particular, by taking ancillary spaces isomorphic to $\HS^{A_IA_O}$ and $\HS^{B_IB_O}$ and appropriately chosen instruments $M_a^{\tilde{A}A},M_b^{\tilde{B}B}$, when Alice and Bob observe $a=b=0$ their operations effectively ``teleport'' $W^{AB}$ to the ancillary spaces so that $E_{0,0}^{\tilde{A}\tilde{B}}$ is (up to normalisation) formally the same as $W^{AB}$. One can then show that if $W^{AB}$ cannot be decomposed as in Eq.~\eqref{eq:csep2}, then the D-POVM element $E_{0,0}^{\tilde{A}\tilde{B}}$ generated in this way can also not be decomposed as in Eq.~\eqref{eq:csep_POVM_00}. Full details of the argument are given in Appendix~\ref{app:MDCI_cert}.

Since matrices of the form of Eq.~\eqref{eq:csep_POVM_00} can be characterised via SDP, one can once again use techniques similar to causal witnesses to certify that a D-POVM is \emph{not} of this form (see Appendix~\ref{app:cones_charact}).
Just as in the SDI-QI scenario, we can then compute the observed witness ``value'' and thereby certify any bipartite causally nonseparable process matrix in an MDCI way, including those that cannot violate causal inequalities.

We note that an analogous result and systematic construction is also known for MDIEWs~\cite{buscemi12,branciard13}.
In contrast to that result, however, the extra MDCI structure assumed in Eq.~\eqref{eq:cstr_instr_v0} is crucial here: in the standard SDI-QI case where it is not assumed, no specific structure is imposed in general on a single D-POVM element generated by a causally separable process matrix (see Appendix~\ref{app:MDCI_cert}), and the SDI-QI certification of the previous section thus required considering the full D-POVM.

\textit{Generalisation to the quantum switch scenario.---}%
A causally nonseparable process that has received significant interest is the ``quantum switch''~\cite{chiribella13}, a tripartite process in which the order of Alice and Bob's operations on some ``target system'' is coherently controlled by the state of a ``control qubit'', given to a third party, Fiona, at the end.
The quantum switch provides advantages in several tasks~\cite{chiribella12,araujo14,guerin16} and, unlike any known bipartite causally nonseparable process, has a clear physical interpretation.
Indeed, several experimental realisations have been performed~\cite{procopio15,rubino17,goswami18,wei19,guo20}.

The quantum switch can be described as a process matrix $W_\text{QS} \in\L(\HS^{ABF})$ in a restricted tripartite scenario---which we call the \textit{``(2+$F$)-partite scenario''}---in which Fiona has no output Hilbert space and simply performs a measurement.
In this scenario, the only relevant causal orders are $A\prec B\prec F$ and $B\prec A\prec F$~\cite{araujo15}, and the generalisation of Eq.~\eqref{eq:gen_Born}, as well as the definitions of causally separable process matrices and D-POVMs is straightforward (see Appendix~\ref{app:2Fcase} for details).
$W_\text{QS}$ is known to be causally nonseparable but to only generate causal correlations~\cite{araujo15,oreshkov16}.
Its importance as a resource in many tasks makes certifying its causal nonseparability a key problem, and multiple experiments have done this in a DD way~\cite{rubino17,goswami18}.

Can this important process be certified in an SDI-QI or MDCI way, despite being extensibly causal~\cite{oreshkov16}?
We find that in both scenarios the response is positive.
Indeed, in the SDI-QI case (i.e., without assuming any structure on the instruments used) and taking a qubit target system and qubit ancillary systems (quantum inputs) for Alice and Bob, and without any quantum input for Fiona, the instruments
\begin{align}
\begin{array}{rl}
     M_a^{\tilde{A}A} & \!= \ketbra{a}{a}^{A_I}\otimes\dketbra{\id}{\id}^{\tilde{A}A_O}\!, \\
     M_b^{\tilde{B}B} & \!= \ketbra{b}{b}^{B_I}\,\otimes\dketbra{\id}{\id}^{\tilde{B}B_O}\!,
\end{array}
\quad M_\pm^{F} = \ketbra{\pm}{\pm}^{F}\!, \label{eq:Ms_QS}
\end{align}
with $\dket{\id}^{\tilde{A}A_O} = \sum_i \ket{i}^{\tilde{A}}\otimes\ket{i}^{A_O}$ and similarly for $\dket{\id}^{\tilde{B}B_O}$ (cf.\ Appendix~\ref{app:Choi_link_prod})
give a causally nonseparable D-POVM (see Appendix~\ref{app:QSwitch}).
These instruments can be interpreted as Alice and Bob performing computational basis measurements on the untrusted systems they receive from the process (in $\HS^{A_I}$ and $\HS^{B_I}$, resp.) while sending their quantum inputs to the process via identity channels; Fiona then measures in the basis $\{\ket{\pm}=\frac{1}{\sqrt{2}}(\ket{0}\pm\ket{1})\}_\pm$.

To understand how robust this certification is, we can consider the robustness of causal nonseparability to noise.
Let us consider the ``depolarised'' quantum switch
\begin{align}
W_\text{QS}(r) = {\textstyle \frac{1}{1+r}}(W_\text{QS} + r \, \id^{ABF}\!/8) \label{eq:WQSr}
\end{align}
parameterised by $r\ge 0$;
it is known that $W_\text{QS}(r)$ is causally nonseparable for $r \lesssim 1.576$~\cite{branciard16a}. 
With the instruments~\eqref{eq:Ms_QS} it is readily checked that $W_\text{QS}(r)$ generates a causally nonseparable D-POVM for $r \lesssim 0.367$ (see Appendix~\ref{app:QSwitch}). 
Despite extensive numerical searches, we were unable to find instruments allowing us to certify the causal nonseparability of $W_\text{QS}(r)$ for $0.367 \lesssim r \lesssim 1.576$ with our SDI-QI approach.
It thus seems that this approach cannot certify all causally nonseparable processes (we found a similar ``robustness gap'' for the bipartite process of Ref.~\cite{feix16} discussed above), in contrast to the MDI certification of entanglement and the MDCI certification of causal nonseparability in the bipartite case.
Nevertheless, the fact our approach provides a noise robust SDI-QI certification of the quantum switch is of significant relevance, given that it is responsible for most known applications of causal nonseparability and yet cannot be certified in a fully DI manner.

One may wonder whether the bipartite results on MDCI witnesses generalise straightforwardly to the (2+$F$)-partite case.
Surprisingly, this turns out not to be the case.
Nonetheless, one can show that MDCI certification is possible for some important classes of processes in this scenario: the ``TTU-'' and ``TUU-noncausal'' processes of Ref.~\cite{bavaresco19}.
These include, in particular, the depolarised quantum switch $W_\text{QS}(r)$ of Eq.~\eqref{eq:WQSr} for $r \lesssim 1.319$~\cite{bavaresco19}, significantly improving the noise tolerance obtained above for SDI-QI certification without the additional MDCI assumption, showing how robustly the quantum switch can be certified with only rather weak assumptions about the performed operations. 
Nonetheless, there remains a gap for $1.319\lesssim r\lesssim 1.576$ where it is open whether $W_\text{QS}(r)$ can be certified in a MDCI way.
A detailed study and discussion of this is given in Appendix~\ref{app:MDCI_cert}.

\textit{Discussion.---}%
In this contribution we significantly relaxed the assumptions required to certify the causal nonseparability of many processes, investigating both SDI-QI and MDCI scenarios.
Notably, we showed how the quantum switch can be certified in an SDI-QI way, and that \emph{all} bipartite causally nonseparable process matrices can be certified in an MDCI manner.

One key open question is to understand precisely which causally nonseparable processes can be certified in an SDI-QI way.
Our inability to find instruments generating a causally nonseparable D-POVM from $W_\text{QS}(r)$ for $0.367 \lesssim r \lesssim 1.576$ indeed leads us to conjecture that \emph{some} such processes cannot be certified in this way.

Beyond understanding fully the bipartite case, an important future direction is the generalisation to multipartite process matrices, where the definition of causal (non)separability is more subtle~\cite{oreshkov16,wechs19}.
One may wonder, for example, whether one can provide an SDI-QI or MDCI certification of more general quantum circuits with quantum control of causal order than just the quantum switch, which can also not violate causal inequalities~\cite{wechs21}.
Another interesting direction is whether our SDI-QI approach can be combined with self-testing techniques to construct fully DI witnesses (as, e.g., in Refs.~\cite{bowles18,supic20} for the case of entanglement).
More broadly, we believe that the notion of causally nonseparable D-POVMs we introduced may be of independent interest to study in its own right; this also suggests that new types of causal nonseparability could be defined, for other kinds of objects beyond process matrices and D-POVMs.
Finally, the idea of imposing extra structure on the instruments used (as in the MDCI scenario) could be adapted to a wide range of quantum resources, opening up new approaches for their certification and exploitation.

\textit{Acknowledgements.---}
We thank Marco Túlio Quintino for enlightening discussions, and acknowledge financial support from the Swiss National Science Foundation (NCCR SwissMAP).

\clearpage
\appendix

\section{Choi isomorphism and link product}
\label{app:Choi_link_prod}

\subsection{Choi isomorphism}

The process matrix formalism relies on the Choi (or Choi-Jamio{\l}kowski) isomorphism~\cite{jamiolkowski72,choi75} to describe quantum operations as matrices. Different versions of the isomorphism can be found in the literature, but in this paper we use the following one (which, in particular, differs from that originally used in Ref.~\cite{oreshkov12} by a transpose): for a given linear map $\M: \L(\HS^X)\to\L(\HS^Y)$, we define its Choi matrix as
\begin{align}
M^{XY} \coloneqq \ & (\I^X \otimes \M)(\dketbra{\id}{\id}^X) \notag \\
= \ & \sum_{i,i'} \ketbra{i}{i'}^X \otimes \M(\ketbra{i}{i'}^X) \quad \in \L(\HS^{XY}),
\end{align}
where $\I^X$ is the identity map on $\L(\HS^X)$, $\dket{\id}^X \coloneqq \sum_i \ket{i}^X \otimes \ket{i}^X$ is a (nonnormalised) maximally entangled state and $\{\ket{i}^X\}_i$ is a fixed (so-called ``computational'') orthonormal basis of $\HS^X$.%
\footnote{When considering isomorphic Hilbert spaces $\HS^X$ and $\HS^{X'}$, we take their computational bases $\{\ket{i}^{X^{(\prime)}}\}_i$ to be in one-to-one correspondence, which allows us to also define $\dket{\id}^{XX'} \coloneqq \sum_i \ket{i}^X \otimes \ket{i}^{X'}$---as we wrote, e.g., in Eq.~\eqref{eq:Ms_QS}.}

Let us recall that a linear map $\M: \L(\HS^X)\to\L(\HS^Y)$ is completely positive if and only if (iff) its Choi matrix $M^{XY} \in \L(\HS^{XY})$ is positive semidefinite (PSD), and that it is trace-preserving (TP) iff its Choi matrix satisfies $\Tr_Y M^{XY} = \id^X$ (where $\Tr_Y$ denotes the partial trace over $\HS^Y$).
For simplicity we directly identify, throughout the paper, linear maps with their Choi matrices.

\subsection{Link product}

The link product was originally introduced in Refs.~\cite{chiribella08,chiribella09} to describe the composition of linear maps in the Choi matrix representation. 
Consider two composite Hilbert spaces $\HS^{XY} = \HS^X\otimes\HS^Y$ and $\HS^{YZ} = \HS^Y\otimes\HS^Z$ that share the same (possibly trivial, i.e., one-dimensional) space factor $\HS^Y$, while $\HS^X$ and $\HS^Z$ do not overlap. The link product of two matrices $M^{XY} \in \L(\HS^{XY})$ and $N^{YZ} \in \L(\HS^{YZ})$ is  then defined as~\cite{chiribella08,chiribella09,wechs21}
\begin{align}
M^{XY}*N^{YZ} \coloneqq \ & \Tr_Y[(M^{XY}\otimes\id^Z)^{T_Y}(\id^X\otimes N^{YZ})] \notag \\
= \ & (\id^{XZ} \!\otimes\! \dbra{\id}^Y\!) (M^{XY} \!\!\otimes\! N^{YZ}) (\id^{XZ} \!\otimes\! \dket{\id}^Y\!) \notag \\
& \qquad \in \L(\HS^{XZ}), \label{eq:def_link_prod}
\end{align}
where $T_Y$ is the partial transpose over $\HS^Y$ (defined in its computational basis).

The link product is commutative (up to a reordering of the Hilbert spaces) and associative (provided each Hilbert space involved in a multiple link product appears at most twice in all factors; we make sure this is indeed always the case). 
Note that it simplifies to a full trace%
\footnote{In particular, the well-known Born rule that gives the probability for a specific outcome $j$ of a POVM $(E_j)_j$ measured on a state $\rho$ can be written, in terms of the link product, as $P(j) = E_j * \rho = \Tr[E_j^T\rho]$---as, e.g., in Eq.~\eqref{eq:Pab_rhoxrhoy}. Note that this way of applying the POVM elements to $\rho$ differs from the more standard (but equivalent) way by a transpose; throughout this paper we choose this convention for simplicity and consistency with our use of the link product.}
$M^Y*N^Y \coloneqq \Tr[(M^Y)^T N^Y]$ when $\HS^X$ and $\HS^Z$ are trivial, and to a mere tensor product $M^X*N^Z \coloneqq M^X\otimes N^Z$ when $\HS^Y$ is trivial. 
It is also useful to note that $M^{XY} * \id^Y = \Tr_Y M^{XY}$, and that the link product of two PSD matrices is also PSD (or a nonnegative scalar for trivial $\HS^X$ and $\HS^Z$).

\section{Validity constraints for process matrices and D-POVMs}
\label{app:validity_cstrs}

For ease of reference, let us recall here the constraints that a bipartite or (2+$F$)-partite matrix $W$ must satisfy in order to be a valid process matrix. We will then show explicitly that these impose that the D-POVMs introduced in Eq.~\eqref{eq:Pab_rhoxrhoy} of the main text are indeed valid POVMs.

Note that, when there is no possible confusion, we often use the terms process matrix and process interchangeably, both in the main text and appendices.

\subsection{Validity of process matrices}
\label{app:validity_cstrs_W}

We refer the reader to Refs.~\cite{oreshkov12,araujo15} for the derivation of the validity constraints of process matrices, which follow from the requirement that the generalised Born rule of Eq.~\eqref{eq:gen_Born} in the main text must always give valid (i.e., nonnegative and normalised) probabilities.

A convenient way to write these constraints is via the ``trace-out-and-replace'' notation introduced in Ref.~\cite{araujo15}, defined (for some matrix $M \in \L(\HS^{XY})$) as
\begin{align}\label{eq:trade_and_pad}
		{}_{X}M \coloneqq (\Tr_X M) \otimes \frac{\id^X}{d_X}\,, \quad {}_{[1-X]}M \coloneqq M - {}_{X}M,
\end{align}
where $d_X$ generically denotes the dimension of $\HS^X$ (and where the definitions above can be applied recursively, e.g., as in ${}_{[1-X][1-Y]}M = {}_{[1-X]}\big({}_{[1-Y]}M\big) = M - {}_{X}M - {}_{Y}M + {}_{XY}M$).

With this notation, it can be proven~\cite{oreshkov12,araujo15,oreshkov16} that a bipartite matrix $W^{AB} \in \L(\HS^{AB}) = \L(\HS^{A_IA_OB_IB_O})$ is a valid process matrix iff $W^{AB}\ge 0$ (i.e., is PSD), $\Tr W^{AB} = d_{A_OB_O}$ and
\begin{align}
{}_{[1-A_O]B}W^{AB} &= {}_{[1-B_O]A}W^{AB}  \notag \\
&= {}_{[1-A_O][1-B_O]}W^{AB}  = 0. \label{eq:validity_cstr_W}
\end{align}
Similarly in the (2+$F$)-partite case (where a third party is introduced, with some input Hilbert space $\HS^F$ but no output Hilbert space), a matrix $W^{ABF} \in \L(\HS^{ABF}) = \L(\HS^{A_IA_OB_IB_OF})$ is a valid process matrix iff it is PSD, $\Tr W^{ABF} = d_{A_OB_O}$ and
\begin{align}
{}_{[1-A_O]BF}W^{ABF} &= {}_{[1-B_O]AF}W^{ABF}  \notag \\
&= {}_{[1-A_O][1-B_O]F}W^{ABF}  = 0. \label{eq:validity_cstr_W_2F}
\end{align}

\medskip

All $W$ matrices considered in the present paper are (at least implicitly) assumed to be valid.
In particular, in the definition of bipartite causal (non)separability, when writing that a causally nonseparable process matrix $W^{AB}$ \emph{cannot} be decomposed as
\begin{align}
q\,W^{A\prec B_I} \otimes \id^{B_O} + (1{-}q)\,W^{B\prec A_I} \otimes \id^{A_O}, \label{eq:csep2_app}
\end{align}
with $W^{A\prec B_I}, W^{B\prec A_I} \ge 0$, it is implicitly required that $W^{A\prec B_I}$ and $W^{B\prec A_I}$ are also valid process matrices, and hence that they also satisfy (according to Eq.~\eqref{eq:validity_cstr_W})
\begin{align}
{}_{[1-A_O]B_I}W^{A\prec B_I} = 0 \ \ \text{and} \ \ {}_{[1-B_O]A_I}W^{B\prec A_I} = 0, \label{eq:constr_W_AprecBI}
\end{align}
respectively. 
As it turns out, however, even if we do not impose Eq.~\eqref{eq:constr_W_AprecBI} \emph{a priori}, a bipartite causally nonseparable process matrix can not be decomposed as in Eq.~\eqref{eq:csep2_app} above: indeed the former would in fact be implied anyway by the validity of $W^{AB}$, via the constraints of Eq.~\eqref{eq:validity_cstr_W}, and the decomposition of Eq.~\eqref{eq:csep2_app}.%
\footnote{This can be seen by writing $q \, \big( {}_{[1-A_O]B_I} W^{A\prec B_I} \big)\otimes \id^{B_O} = {}_{[1-A_O]B} \big( q\, W^{A\prec B_I}\otimes \id^{B_O} \big) = {}_{[1-A_O]B} [W^{AB} - (1{-}q)\,W^{B\prec A_I} \otimes \id^{A_O}] = {}_{[1-A_O]B} W^{AB} - (1{-}q)\,{}_{[1-A_O]B} \big(W^{B\prec A_I} \otimes \id^{A_O} \big) = 0$, which indeed implies (for $q>0$) that ${}_{[1-A_O]B_I} W^{A\prec B_I} = 0$; and similarly for ${}_{[1-B_O]A_I}W^{B\prec A_I} = 0$. \label{footnote:valid_a_fortiori}}

\subsection{Validity of the induced D-POVMs}
\label{app:validity_induced_povm}

For the sake of completeness, and as a sanity check, let us show explicitly that the set of operators
\begin{align}
E_{a,b}^{\tilde{A}\tilde{B}} = \left(M_a^{\tilde{A}A}\otimes M_b^{\tilde{B}B}\right)*W^{AB}
\label{eq:Eab}
\end{align}
introduced in Eq.~\eqref{eq:Pab_rhoxrhoy} defines a valid POVM. As we will see, this indeed follows from the validity constraints of the instruments $(M_a^{\tilde{A}A})_a$, $(M_b^{\tilde{B}B})_b$ and of the process matrix $W^{AB}$.

First note that since all $M_a^{\tilde{A}A}, M_b^{\tilde{B}B}, W^{AB}$'s are PSD, then it clearly follows from Eq.~\eqref{eq:Eab} that all $E_{a,b}^{\tilde{A}\tilde{B}}$'s are PSD as well.

It remains to be verified that these sum up to the identity. Using ${}_{[1-A_O][1-B_O]}W^{AB} = 0$ from Eq.~\eqref{eq:validity_cstr_W}, we can write
\begin{align}
\sum_{a,b} E_{a,b}^{\tilde{A}\tilde{B}} = & \big(\sum_aM_a^{\tilde{A}A}\otimes\sum_b M_b^{\tilde{B}B}\big)*W^{AB}\notag \\
= & \big(\sum_aM_a^{\tilde{A}A}\otimes\sum_b M_b^{\tilde{B}B}\big)*{}_{B_O}W^{AB}\notag \\
& + \big(\sum_aM_a^{\tilde{A}A}\otimes\sum_b M_b^{\tilde{B}B}\big)*{}_{A_O}W^{AB}\notag \\
& - \big(\sum_a\!M_a^{\tilde{A}A}\otimes\!\sum_b \!M_b^{\tilde{B}B}\big)\!*\!{}_{A_OB_O\!}W^{AB}. \label{eq:sum_tilde_Wab}
\end{align}
Consider the first term in the last equality; using the properties of the link product and the TP constraint $\Tr_{B_O}\sum_b M_b^{\tilde{B}B} = \id^{\tilde{B}B_I}$, we have
\begin{align}
& \big(\sum_aM_a^{\tilde{A}A}\otimes\sum_b M_b^{\tilde{B}B}\big)*{}_{B_O}W^{AB}\notag \\
& = \big(\sum_aM_a^{\tilde{A}A}\otimes\sum_b M_b^{\tilde{B}B}\big)*(\Tr_{B_O}W^{AB}\otimes\id^{B_O}/d_{B_O})\notag \\
& = \big(\sum_aM_a^{\tilde{A}A}\otimes\Tr_{B_O}\sum_b M_b^{\tilde{B}B}\big)*\Tr_{B_O}W^{AB}/d_{B_O}\notag \\
& = \big(\sum_aM_a^{\tilde{A}A}\otimes\id^{\tilde{B}B_I}\big)*\Tr_{B_O}W^{AB}/d_{B_O}\notag \\
& = \big(\sum_aM_a^{\tilde{A}A}\big)*(\Tr_B W^{AB})\otimes\id^{\tilde{B}}/d_{B_O}.
\end{align}
Using now ${}_{[1-A_O]B}W^{AB} = 0$, i.e., equivalently, $\Tr_B W^{AB} = {}_{A_O}(\Tr_B W^{AB}) = (\Tr_{A_OB}W^{AB}) \otimes \id^{A_O}/d_{A_O}$, and the TP constraint $\Tr_{A_O}\sum_a M_a^{\tilde{A}A} = \id^{\tilde{A}A_I}$, we get
\begin{align}
& \big(\sum_aM_a^{\tilde{A}A}\big)*(\Tr_B W^{AB})\otimes\id^{\tilde{B}}/d_{B_O}\notag \\
& = \big(\sum_aM_a^{\tilde{A}A}\big)*(\Tr_{A_OB}W^{AB}\otimes\id^{A_O})\otimes\id^{\tilde{B}}/(d_{A_O}d_{B_O})\notag \\
& = \big(\Tr_{A_O}\sum_aM_a^{\tilde{A}A}\big)*(\Tr_{A_OB}W^{AB})\otimes\id^{\tilde{B}}/(d_{A_O}d_{B_O})\notag \\
& = \id^{\tilde{A}A_I}*(\Tr_{A_OB}W^{AB})\otimes\id^{\tilde{B}}/(d_{A_O}d_{B_O})\notag \\
& = (\Tr_{AB}W^{AB})\,\id^{\tilde{A}\tilde{B}}/(d_{A_O}d_{B_O}) = \id^{\tilde{A}\tilde{B}}.
\end{align}
Similarly the last 2 terms in Eq.~\eqref{eq:sum_tilde_Wab} above also give $\id^{\tilde{A}\tilde{B}}$, so that we end up with $\sum_{a,b} E_{a,b}^{\tilde{A}\tilde{B}} = \id^{\tilde{A}\tilde{B}}$, as required.

In the (2+$F$)-partite case, noting that $\sum_f M_f^{\tilde{F}F} = \id^{\tilde{F}F}$, one has $\sum_{a,b,f} E_{a,b,f}^{\tilde{A}\tilde{B}\tilde{F}} = \big(\sum_aM_a^{\tilde{A}A}\otimes\sum_b M_b^{\tilde{B}B}\big)*\Tr_F W^{ABF}\otimes\id^{\tilde{F}}$, so that all the calculations above generalise straightforwardly, by replacing $W^{AB}$ by $\Tr_F W^{ABF}\otimes\id^{\tilde{F}}$.

\section{D-POVMs induced by causally ordered process matrices}
\label{app:causallySepDPOVMs}

Here we prove the no-signalling condition on D-POVMs arising from a causally ordered process given by Eq.~\eqref{eq:nosig_E_AprecB} of the main text.

Consider a process matrix $W^{AB} = W^{A\prec B_I}\otimes \id^{B_O}$ compatible with the order $A\prec B$. 
Using the fact that $M^{XY} * \id^Y = \Tr_Y M^{XY}$ and that $\Tr_{B_O} \sum_b M_b^{\tilde{B}B} = \id^{\tilde{B}B_I}$ by the TP condition (see Sec.~\ref{app:Choi_link_prod}), one readily sees that
\begin{align}
\sum_b E_{a,b}^{\tilde{A}\tilde{B}} & ={\textstyle \sum_b} \, (M_a^{\tilde{A}A}\otimes M_b^{\tilde{B}B})*(W^{A\prec B_I}\otimes \id^{B_O}) \notag \\[-2mm]
& = (M_a^{\tilde{A}A}\otimes \Tr_{B_O} {\textstyle \sum_b} \,M_b^{\tilde{B}B})* W^{A\prec B_I} \notag \\[1mm]
& = (M_a^{\tilde{A}A}\otimes \id^{\tilde{B}B_I})* W^{A\prec B_I} \notag\\
& = E_a^{\tilde{A}} \otimes \id^{\tilde{B}}
\end{align}
with $E_a^{\tilde{A}} = M_a^{\tilde{A}A} * \Tr_{B_I} W^{A\prec B_I} \ge 0$ defining a (single-partite) POVM $(E_a^{\tilde{A}})_a$.

Similarly, given a process matrix $W^{AB} = W^{B\prec A_I}\otimes \id^{A_O}$ compatible with the order $B\prec A$, one sees that 
\begin{align}
\sum_a E_{a,b}^{\tilde{A}\tilde{B}} = \id^{\tilde{A}} \otimes  E_b^{\tilde{B}}  \label{eq:nosig_E_BprecA}
\end{align}
with $E_b^{\tilde{B}} = M_b^{\tilde{B}B} * \Tr_{A_I} W^{B\prec A_I} \ge 0$ defining a (single-partite) POVM $(E_b^{\tilde{B}})_b$.

The fact that both $(E_a^{\tilde{A}})_a$ and $(E_b^{\tilde{B}})_b$ are indeed valid POVMs follows immediately from the fact that $\mathbb{E}^{\tilde{A}\tilde{B}}=(E_{a,b}^{\tilde{A}\tilde{B}})_{a,b}$ is a valid POVM, as shown in Sec.~\ref{app:validity_induced_povm} above.

\section{Causal (non)separability in the (2+$F$)-partite case}
\label{app:2Fcase}

The (2+$F$)-partite case corresponds to a particular tripartite scenario with two parties, Alice and Bob, having both some input ($\HS^{A_I}, \HS^{B_I}$) and output ($\HS^{A_O}, \HS^{B_O}$) Hilbert spaces, while the third party, Fiona, has an input Hilbert space ($\HS^F$) but no output Hilbert space.

Although extending the notion of causal (non)separability to the multipartite case is in general not so straightforward~\cite{oreshkov16,wechs19}, its extension to the specific (2+$F$)-partite case remains rather simple~\cite{araujo15}. Indeed, because Fiona has no output Hilbert space, she can always be taken to act last, in the causal future of both Alice and Bob, so that the only relevant causal orders are $A\prec B\prec F$ and $B\prec A\prec F$. The definition of causal (non)separability for process matrices then simply generalises as follows: causally separable process matrices are those of the form
\begin{align}
W^{ABF} & = q\,W^{A\prec B\prec F} + (1{-}q)\,W^{B\prec A\prec F}
\label{eq:csep2F}
\end{align}
with $q\in[0,1]$ and where $W^{A\prec B\prec F}, W^{B\prec A\prec F}\in\L(\HS^{ABF})$ are causally ordered process matrices, such that $\Tr_F W^{A\prec B\prec F} = W^{A\prec B_I} \otimes \id^{B_O}$ and $\Tr_F W^{B\prec A\prec F} = W^{B\prec A_I} \otimes \id^{A_O}$~\cite{araujo15,oreshkov16}.

\medskip

In a scenario with quantum inputs, Eq.~\eqref{eq:Pab_rhoxrhoy} generalises easily to the (2+$F$)-partite case, so that one is led to consider the D-POVM $\mathbb{E}^{\tilde{A}\tilde{B}\tilde{F}} = (E_{a,b,f}^{\tilde{A}\tilde{B}\tilde{F}})_{a,b,f}$ with
\begin{align}
E_{a,b,f}^{\tilde{A}\tilde{B}\tilde{F}} = \left(M_a^{\tilde{A}A}\otimes M_b^{\tilde{B}B}\otimes M_f^{\tilde{F}F}\right)*W^{ABF},
\label{eq:Eabf}
\end{align}
thereby generalising Eq.~\eqref{eq:Eab}.

It is easily seen, in a similar way to the bipartite case (see Sec.~\ref{app:causallySepDPOVMs} above), that if $W^{ABF} = W^{A\prec B\prec F}$ is compatible with the order $A\prec B\prec F$, then the induced D-POVM $\mathbb{E}^{\tilde{A}\tilde{B}\tilde{F}}$ is compatible with the causal order $\tilde{A}\prec\tilde{B}\prec\tilde{F}$, and satisfies $\sum_f E_{a,b,f}^{\tilde{A}\tilde{B}\tilde{F}} = E_{a,b}^{\tilde{A}\prec\tilde{B}} \otimes \id^{\tilde{F}}$ for all $a,b$ and for some bipartite D-POVM $(E_{a,b}^{\tilde{A}\prec\tilde{B}})_{a,b}$ compatible with $\tilde{A}\prec\tilde{B}$ (hence, further satisfying $\sum_b E_{a,b}^{\tilde{A}\prec\tilde{B}} = E_a^{\tilde{A}} \otimes \id^{\tilde{B}}$ for all $a$); we generically denote such a D-POVM $\mathbb{E}^{\tilde{A}\prec\tilde{B}\prec\tilde{F}} = (E_{a,b,f}^{\tilde{A}\prec\tilde{B}\prec\tilde{F}})_{a,b,f}$ (and similarly for the order $\tilde{B}\prec\tilde{A}\prec\tilde{F}$). 

Analogously to the bipartite case, we then define:%
\footnote{Note that the structure of the space in which the D-POVM $\mathbb{E}^{\tilde{A}\tilde{B}\tilde{F}}$ is defined does not reflect the fact that Fiona had no output Hilbert space, and therefore does not by itself imply that $\tilde{F}$ is taken to come last; hence the clarification in Definition~\ref{def:csep_POVM_2F}.}
\setcounter{theorem}{1}
\begin{definition}\label{def:csep_POVM_2F}
A (2+$F$)-partite D-POVM $\mathbb{E}^{\tilde{A}\tilde{B}\tilde{F}}$ (where $\tilde{F}$ comes last) that can be decomposed as a convex mixture of D-POVMs compatible with the causal orders $\tilde{A}\prec\tilde{B}\prec\tilde{F}$ and $\tilde{B}\prec\tilde{A}\prec\tilde{F}$, i.e., of the form
\begin{align}
\mathbb{E}^{\tilde{A}\tilde{B}\tilde{F}} = q\, \mathbb{E}^{\tilde{A}\prec\tilde{B}\prec\tilde{F}} + (1{-}q)\, \mathbb{E}^{\tilde{B}\prec\tilde{A}\prec\tilde{F}} \label{eq:csep_POVM_2F}
\end{align}
with $q\in[0,1]$ is said to be \emph{causally separable}.
\end{definition}

It is again clear that a causally separable process matrix can only generate causally separable D-POVMs. As in the bipartite case, the converse also holds: any causally separable D-POVM can be realised by local operations on a causally separable process matrix (see Sec.~\ref{app:any_csep_DPOVM} below).

\section{Realisation of any causally separable D-POVM}
\label{app:any_csep_DPOVM}

Here we show, in both the bipartite and the (2+$F$)-partite cases, that any causally separable D-POVM can be realised by local operations on a causally separable process matrix. We provide for this some explicit constructions, inspired by those in Appendix~B of Ref.~\cite{wechs21}.

\subsection{In the bipartite case}

We will first show how to realise a D-POVM compatible with a single, fixed order, before showing how to generalise this to causally separable D-POVMs.
We thus begin by considering a D-POVM $\mathbb{E}^{\tilde{A}\prec\tilde{B}} = (E_{a,b}^{\tilde{A}\prec\tilde{B}})_{a,b}$ compatible with the order $\tilde{A}\prec\tilde{B}$, such that $\sum_b E_{a,b}^{\tilde{A}\prec\tilde{B}} = E_a^{\tilde{A}} \otimes \id^{\tilde{B}}$ for all $a$.

Informally, our construction proceeds by extending Alice's POVM $E_a^{\tilde{A}}$ to an appropriate, natural instrument and sending the transformed state, along with the measurement outcome $a$, to Bob via a trivial identity process.
We achieve this by first purifying $E_a$ and using a different purifying ``output'' space for each $a$, so that the state is sent to the process in different subspaces depending on $a$.
We then show how to find a measurement for Bob that gives the correct global D-POVM by ``inverting'' the link product using techniques adapted from Ref.~\cite{wechs21}.

Formally, note that since (for all $a,b$) $E_a^{\tilde{A}}$ and $E_{a,b}^{\tilde{A}\prec\tilde{B}}$ are PSD, these admit spectral decompositions of the form
\begin{align}
E_a^{\tilde{A}} & = \sum_{i_a} \ketbra{e_a^{i_a}}{e_a^{i_a}}^{\tilde{A}}, \notag \\
E_{a,b}^{\tilde{A}\prec\tilde{B}} & = \sum_j \ketbra{e_{a,b}^j}{e_{a,b}^j}^{\tilde{A}\prec\tilde{B}}
\end{align}
for some orthogonal sets of $r_a$ and $r_{a,b}$ (nonnormalised and nonzero) vectors $\{\ket{e_a^{i_a}}^{\tilde{A}}\}_{i_a}$ and $\{\ket{e_{a,b}^j}^{\tilde{A}\prec\tilde{B}}\}_j$, resp.
Let us introduce, for each $a$, some $r_a$-dimensional Hilbert space $\HS^{A_O^{(a)}}$ with computational basis $\{\ket{i_a}^{A_O^{(a)}}\}_{i_a}$, and their direct sum $\HS^{A_O} \coloneqq \bigoplus_{a} \HS^{A_O^{(a)}}$ with computational basis $\{\ket{i_a}^{A_O}\}_{a,i_a}$ (obtained by embedding each $\ket{i_a}^{A_O^{(a)}} \in \HS^{A_O^{(a)}}$ into the larger space $\HS^{A_O}$; we take the $i_a$'s and $i_{a'}$'s for $a\neq a'$ to be different). Let us also introduce some Hilbert spaces $\HS^{B_I^{(a)}}$ and $\HS^{B_I}$ isomorphic to $\HS^{A_O^{(a)}}$ and $\HS^{A_O}$, resp. $\HS^{A_O}$ and $\HS^{B_I}$ define Alice's output and Bob's input spaces; we take Alice's input and Bob's output spaces, on the other hand, to be trivial, and define the (identity channel) process matrix
\begin{align}
W^{A_O\prec B_I} = \dketbra{\id}{\id}^{A_OB_I}.
\end{align}

Let us then define
\begin{align}
\ket{m_a}^{\tilde{A}A_O} & = \sum_{i_a} \ket{e_a^{i_a}}^{\tilde{A}}\otimes\ket{i_a}^{A_O}, \notag \\
\ket{m_{b|a}^j}^{\tilde{B}B_I} & = \sum_{i_a} \Big({\textstyle \frac{\bra{e_a^{i_a}}^{\tilde{A}}}{\braket{e_a^{i_a}}{e_a^{i_a}}^{\tilde{A}}}}\otimes\id^{\tilde{B}}\Big)\ket{e_{a,b}^j}^{\tilde{A}\prec\tilde{B}}\otimes\ket{i_a}^{B_I}, \notag \\
M_a^{\tilde{A}A_O} & = \ketbra{m_a}{m_a}^{\tilde{A}A_O} \quad (\ge 0), \notag \\
M_b^{\tilde{B}B_I} & = \sum_{j,a} \ketbra{m_{b|a}^j}{m_{b|a}^j}^{\tilde{B}B_I} \quad (\ge 0).
\end{align}
Note that $\ket{m_a}^{\tilde{A}A_O}$ is such that $\Tr_{A_O} \ketbra{m_a}{m_a}^{\tilde{A}A_O} = E_a^{\tilde{A}}$, which implies in particular that $(M_a^{\tilde{A}A_O})_a$ is a valid instrument. Using $\sum_{j,b} \ketbra{e_{a,b}^j}{e_{a,b}^j}^{\tilde{A}\prec\tilde{B}} = \sum_b E_{a,b}^{\tilde{A}\prec\tilde{B}} = E_a^{\tilde{A}} \otimes \id^{\tilde{B}}$ and $\frac{\bra{e_a^{i_a}}^{\tilde{A}}E_a^{\tilde{A}}\ket{e_a^{i_a'}}^{\tilde{A}}}{\braket{e_a^{i_a}}{e_a^{i_a}}^{\tilde{A}}\braket{e_a^{i_a'}}{e_a^{i_a'}}^{\tilde{A}}} = \delta_{i_a,i_a'}$ (where $\delta$ is the Kronecker delta), we note also that $\sum_b M_b^{\tilde{B}B_I} = \id^{\tilde{B}B_I}$, so that $(M_b^{\tilde{B}B_I})_b$ is a valid POVM.

With these definitions, one obtains%
\footnote{Here we use the link product for vectors, defined for $\ket{m}^{XY}\in\HS^{XY}$ and $\ket{n}^{YZ}\in\HS^{YZ}$ as $\ket{m}^{XY}*\ket{n}^{YZ}\coloneqq (\id^{XZ} \otimes \dbra{\id}^Y) (\ket{m}^{XY}\otimes\ket{n}^{YZ}) \in \HS^{XZ}$~\cite{wechs21}. It is such that $(\ket{m}^{XY}*\ket{n}^{YZ})(\bra{m}^{XY}*\bra{n}^{YZ}) = \ketbra{m}{m}^{XY}*\ketbra{n}{n}^{YZ}$, cf.\ Eq.~\eqref{eq:def_link_prod}, as we use in Eq.~\eqref{eq:Ma_Mb_W}.}
\begin{align}
& \big(\ket{m_a}^{\tilde{A}A_O}\otimes \ket{m_{b|a'}^j}^{\tilde{B}B_I}\big) * \dket{\id}^{A_OB_I} \notag \\
& = \delta_{a,a'} \sum_{i_a} \Big({\textstyle \frac{\ketbra{e_a^{i_a}}{e_a^{i_a}}^{\tilde{A}}}{\braket{e_a^{i_a}}{e_a^{i_a}}^{\tilde{A}}}}\otimes\id^{\tilde{B}}\Big)\ket{e_{a,b}^j}^{\tilde{A}\prec\tilde{B}} = \delta_{a,a'} \ket{e_{a,b}^j}^{\tilde{A}\prec\tilde{B}}, \label{eq:ma_mba_Id}
\end{align}
where we used the facts that $\sum_{i_a} \frac{\ketbra{e_a^{i_a}}{e_a^{i_a}}^{\tilde{A}}}{\braket{e_a^{i_a}}{e_a^{i_a}}^{\tilde{A}}}$ acts as the identity on $\range(E_a^{\tilde{A}})$ and that $\ket{e_{a,b}^j}^{\tilde{A}\prec\tilde{B}} \in \range(E_a^{\tilde{A}}) \otimes \HS^{\tilde{B}}$.%
\footnote{This can, e.g., be seen by contradiction: suppose that $\ket{e_{a,b}^j}^{\tilde{A}\prec\tilde{B}} \notin \range(E_a^{\tilde{A}}) \otimes \HS^{\tilde{B}}$. Then $\exists \, \ket{v}\in\range(E_a^{\tilde{A}})^\perp$ s.t. $(\bra{v}^{\tilde{A}}\otimes \id^{\tilde{B}}) \ket{e_{a,b}^j}^{\tilde{A}\prec\tilde{B}} \neq 0$. This implies (since all terms are PSD) $\sum_{j,b} (\bra{v}\otimes \id) \ketbra{e_{a,b}^j}{e_{a,b}^j}^{\tilde{A}\prec\tilde{B}} (\ket{v}\otimes \id) = \sum_b (\bra{v}\otimes \id) E_{a,b}^{\tilde{A}\prec\tilde{B}} (\ket{v}\otimes \id) = (\bra{v}\otimes \id) E_a^{\tilde{A}} \otimes \id^{\tilde{B}} (\ket{v}\otimes \id) = \bra{v} E_a^{\tilde{A}} \ket{v}\, \id^{\tilde{B}} \neq 0$, in contradiction with $\ket{v}\in\range(E_a^{\tilde{A}})^\perp$.}
We thus find
\begin{align}
& (M_a^{\tilde{A}A_O}\otimes M_b^{\tilde{B}B_I}) * W^{A_O\prec B_I} \notag \\
& = (\ketbra{m_a}{m_a}^{\tilde{A}A_O}\otimes {\textstyle \sum_{j,a'}} \ketbra{m_{b|a'}^j}{m_{b|a'}^j}^{\tilde{B}B_I}) * \dketbra{\id}{\id}^{A_OB_I} \notag \\
& = \sum_j \ketbra{e_{a,b}^j}{e_{a,b}^j}^{\tilde{A}\prec\tilde{B}} = E_{a,b}^{\tilde{A}\prec\tilde{B}}, \label{eq:Ma_Mb_W}
\end{align}
so that our choice of process and of instruments above indeed allowed us to generate the causally ordered D-POVM $\mathbb{E}^{\tilde{A}\prec\tilde{B}} = (E_{a,b}^{\tilde{A}\prec\tilde{B}})_{a,b}$.

\medskip

Consider now a causally separable D-POVM $\mathbb{E} = q\, \mathbb{E}^{\tilde{A}\prec\tilde{B}} + (1{-}q)\, \mathbb{E}^{\tilde{B}\prec\tilde{A}}$.

Using the previous construction, one can obtain some instruments $(M_a^{\tilde{A}A_O\,[\tilde{A}\prec\tilde{B}]})_a$ and $(M_b^{B_I\tilde{B}\,[\tilde{A}\prec\tilde{B}]})_b$ and some process matrix $W^{A_O\prec B_I}$ such that $(M_a^{\tilde{A}A_O\,[\tilde{A}\prec\tilde{B}]} \otimes M_b^{B_I\tilde{B}\,[\tilde{A}\prec\tilde{B}]}) * W^{A_O\prec B_I} = E_{a,b}^{\tilde{A}\prec\tilde{B}}$ for all $a,b$.
With a similar construction (and introducing the appropriate spaces $\HS^{A_I}$ and $\HS^{B_O}$), one can obtain some instruments $(M_a^{A_I\tilde{A}\,[\tilde{B}\prec\tilde{A}]})_a$ and $(M_b^{\tilde{B}B_O\,[\tilde{B}\prec\tilde{A}]})_b$ and some process matrix $W^{B_O\prec A_I}$ such that $(M_a^{A_I\tilde{A}\,[\tilde{B}\prec\tilde{A}]} \otimes M_b^{\tilde{B}B_O\,[\tilde{B}\prec\tilde{A}]}) * W^{B_O\prec A_I} = E_{a,b}^{\tilde{B}\prec\tilde{A}}$ for all $a,b$.
We shall rename the spaces $\HS^{B_I}$ and $\HS^{A_I}$ introduced in these two cases as $\HS^{B_I^0}$ and $\HS^{A_I^0}$, resp.

From these, we now construct some new instruments $(M_a^{\tilde{A}A})_a$, $(M_b^{\tilde{B}B})_b$ and some process matrix $W^{AB}$ as follows.
Let us introduce some qubit (2-dimensional) Hilbert spaces $\HS^\alpha$ and $\HS^\beta$ (used to encode a classical control of the causal order), define $\HS^{A_I} = \HS^{\alpha A_I^0}$, $\HS^{B_I} = \HS^{\beta B_I^0}$, and
\begin{align}
M_a^{\tilde{A}A} = & \ketbra{0}{0}^\alpha \otimes \id^{A_I^0} \otimes M_a^{\tilde{A}A_O\,[\tilde{A}\prec\tilde{B}]} \notag \\
& +  \ketbra{1}{1}^\alpha \otimes M_a^{A_I^0\tilde{A}\,[\tilde{B}\prec\tilde{A}]} \otimes \id^{A_O}/d_{A_O}, \notag \\[1mm]
M_b^{\tilde{B}B} = & \ketbra{0}{0}^\beta \otimes M_b^{B_I^0\tilde{B}\,[\tilde{A}\prec\tilde{B}]} \otimes \id^{B_O}/d_{B_O} \notag \\
& +  \ketbra{1}{1}^\beta \otimes \id^{B_I^0} \otimes M_b^{\tilde{B}B_O\,[\tilde{B}\prec\tilde{A}]}, \notag \\[1mm]
W^{AB} = & q \, \ketbra{0}{0}^\alpha \otimes \ketbra{0}{0}^\beta \otimes \id^{A_I^0}/d_{A_I^0} \otimes W^{A_O\prec B_I^0} \otimes \id^{B_O} \notag \\
& \!+\! (1{-}q) \ketbra{1}{1}^\alpha \!\otimes\! \ketbra{1}{1}^\beta \!\otimes\! \id^{B_I^0}\!/\!d_{B_I^0} \!\otimes\! W^{B_O\prec A_I^0} \!\otimes\! \id^{A_O}.
\end{align}
One can verify that these indeed define valid instruments and a valid causally separable process matrix.

With these definitions we then get
\begin{align}
& (M_a^{\tilde{A}A}\otimes M_b^{\tilde{B}B}) * W^{AB} \notag \\[1mm]
& = q \, (M_a^{\tilde{A}A_O\,[\tilde{A}\prec\tilde{B}]} \otimes M_b^{B_I^0\tilde{B}\,[\tilde{A}\prec\tilde{B}]}) * W^{A_O\prec B_I^0} \notag \\
& \quad + (1{-}q)\, (M_a^{A_I^0\tilde{A}\,[\tilde{B}\prec\tilde{A}]} \otimes M_b^{\tilde{B}B_O\,[\tilde{B}\prec\tilde{A}]}) * W^{B_O\prec A_I^0}  \notag \\[1mm]
& = q \, E_{a,b}^{\tilde{A}\prec\tilde{B}} + (1{-}q)\, E_{a,b}^{\tilde{B}\prec\tilde{A}},
\end{align}
as desired.

\subsection{In the (2+$F$)-partite case}

The previous construction extends easily to the (2+$F$)-partite case. Let us just briefly sketch how this generalisation works.

Similarly to the bipartite case, for a causally ordered D-POVM $\mathbb{E}^{\tilde{A}\prec\tilde{B}\prec\tilde{F}} = (E_{a,b,f}^{\tilde{A}\prec\tilde{B}\prec\tilde{F}})_{a,b,f}$ such that $\sum_f E_{a,b,f}^{\tilde{A}\prec\tilde{B}\prec\tilde{F}} = E_{a,b}^{\tilde{A}\prec\tilde{B}} \otimes \id^{\tilde{F}}$ for all $a,b$ and $\sum_b E_{a,b}^{\tilde{A}\prec\tilde{B}} = E_a^{\tilde{A}} \otimes \id^{\tilde{B}}$ for all $a$, one can introduce the spectral decompositions
\begin{align}
E_a^{\tilde{A}} & = \sum_{i_a} \ketbra{e_a^{i_a}}{e_a^{i_a}}^{\tilde{A}}, \notag \\
E_{a,b}^{\tilde{A}\prec\tilde{B}} & = \sum_{j_{a,b}} \ketbra{e_{a,b}^{j_{a,b}}}{e_{a,b}^{j_{a,b}}}^{\tilde{A}\prec\tilde{B}}, \notag \\
E_{a,b,f}^{\tilde{A}\prec\tilde{B}\prec\tilde{F}} & = \sum_k \ketbra{e_{a,b,f}^k}{e_{a,b,f}^k}^{\tilde{A}\prec\tilde{B}\prec\tilde{F}}.
\end{align}
We then introduce the Hilbert spaces $\HS^{A_O^{(a)}}$ and their direct sum $\HS^{A_O} \coloneqq \bigoplus_{a} \HS^{A_O^{(a)}}$ as before, together now with some spaces $\HS^{B_O^{(a,b)}}$ with computational bases $\{\ket{j_{a,b}}^{B_O^{(a,b)}}\}_{j_{a,b}}$ and their direct sum $\HS^{B_O} \coloneqq \bigoplus_{a,b} \HS^{B_O^{(a,b)}}$ with computational basis $\{\ket{j_{a,b}}^{B_O}\}_{a,b,j_{a,b}}$. We similarly introduce the Hilbert spaces $\HS^{B_I^{(a)}}$ and $\HS^{B_I}$ isomorphic to $\HS^{A_O^{(a)}}$ and $\HS^{A_O}$, as well now as the spaces $\HS^{F^{(a,b)}}$ and $\HS^F$ isomorphic to $\HS^{B_O^{(a,b)}}$ and $\HS^{B_O}$, resp.; and we define the process matrix (with trivial $\HS^{A_I}$)
\begin{align}
W^{A_O\prec B\prec F} = \dketbra{\id}{\id}^{A_OB_I}\otimes \dketbra{\id}{\id}^{B_OF}.
\end{align}

We then define (omitting some tensor products)
\begin{align}
\ket{m_a}^{\tilde{A}A_O} & = \sum_{i_a} \ket{e_a^{i_a}}^{\tilde{A}}\otimes\ket{i_a}^{A_O}, \notag \\
\ket{m_{b|a}}^{\tilde{B}B} \!& = \!\sum_{i_a,j_{a,b}}\!\!\! \Big({\textstyle \frac{\bra{e_a^{i_a}}^{\tilde{A}}}{\braket{e_a^{i_a}}{e_a^{i_a}}^{\tilde{A}}}}\!\otimes\!\id^{\tilde{B}}\Big)\!\ket{e_{a,b}^{j_{a,b}}}^{\!\tilde{A}\prec\tilde{B}}\!\ket{i_a}^{\!B_I}\!\ket{j_{a,b}}^{\!B_O}\!\!, \notag \\
\ket{m_{f|a,b}^k}^{\tilde{F}F} \!& = \sum_{j_{a,b}}\! \Big(\!{\textstyle \frac{\bra{e_{a,b}^{j_{a,b}}}^{\tilde{A}\prec\tilde{B}}}{\braket{e_{a,b}^{j_{a,b}}}{e_{a,b}^{j_{a,b}}}^{\tilde{A}\prec\tilde{B}}}}\!\otimes\!\id^{\tilde{F}}\Big)\!\ket{e_{a,b,f}^k}^{\!\tilde{A}\prec\tilde{B}\prec\tilde{F}}\!\ket{j_{a,b}}^{\!F}\!\!\!, \notag
\end{align}
\begin{align}
M_a^{\tilde{A}A_O} & = \ketbra{m_a}{m_a}^{\tilde{A}A_O} \quad (\ge 0), \notag \\
M_b^{\tilde{B}B} & = \sum_a \ketbra{m_{b|a}}{m_{b|a}}^{\tilde{B}B} \quad (\ge 0), \notag \\
M_f^{\tilde{F}F} & = \sum_{k,a,b} \ketbra{m_{f|a,b}^k}{m_{f|a,b}^k}^{\tilde{F}F} \quad (\ge 0).
\end{align}
One can verify that these define valid instruments (or POVM, for $(M_f^{\tilde{F}F})_f$).

With these definitions, one obtains (similarly to Eqs.~\eqref{eq:ma_mba_Id}--\eqref{eq:Ma_Mb_W})
\begin{align}
& \big(\ket{m_a}^{\tilde{A}A_O}\!\otimes\! \ket{m_{b|a'}}^{\tilde{B}B}\!\otimes\! \ket{m_{f|a'',b'}^k}^{\tilde{F}F}\big) * \dket{\id}^{A_OB_I}\dket{\id}^{B_OF} \notag \\[1mm]
& \ = \delta_{a,a'}\, \delta_{a',a''}\, \delta_{b,b'} \sum_{j_{a,b}} \Big({\textstyle \sum_{i_a} \frac{\ketbra{e_a^{i_a}}{e_a^{i_a}}^{\tilde{A}}}{\braket{e_a^{i_a}}{e_a^{i_a}}^{\tilde{A}}}}\!\otimes\!\id^{\tilde{B}}\Big)\!\ket{e_{a,b}^{j_{a,b}}}^{\!\tilde{A}\prec\tilde{B}} \notag \\[-2mm]
& \hspace{33mm} \Big(\!{\textstyle \frac{\bra{e_{a,b}^{j_{a,b}}}^{\tilde{A}\prec\tilde{B}}}{\braket{e_{a,b}^{j_{a,b}}}{e_{a,b}^{j_{a,b}}}^{\tilde{A}\prec\tilde{B}}}}\!\otimes\!\id^{\tilde{F}}\Big)\!\ket{e_{a,b,f}^k}^{\!\tilde{A}\prec\tilde{B}\prec\tilde{F}} \notag \\[1mm]
& \ = \delta_{a,a'}\, \delta_{a',a''}\, \delta_{b,b'}\! \sum_{j_{a,b}} \!\Big({\textstyle \frac{\ketbra{e_{a,b}^{j_{a,b}}}{e_{a,b}^{j_{a,b}}}^{\tilde{A}\prec\tilde{B}}}{\braket{e_{a,b}^{j_{a,b}}}{e_{a,b}^{j_{a,b}}}^{\tilde{A}\prec\tilde{B}}}}\!\otimes\!\id^{\tilde{F}}\Big)\!\ket{e_{a,b,f}^k}^{\!\tilde{A}\prec\tilde{B}\prec\tilde{F}} \notag \\[1mm]
& \ = \delta_{a,a'}\, \delta_{a',a''}\, \delta_{b,b'} \ket{e_{a,b,f}^k}^{\tilde{A}\prec\tilde{B}\prec\tilde{F}},
\end{align}
and
\begin{align}
& (M_a^{\tilde{A}A_O}\otimes M_b^{\tilde{B}B}\otimes M_f^{\tilde{F}F}) * W^{A_O\prec B\prec F} \notag \\
& = \sum_k \ketbra{e_{a,b,f}^k}{e_{a,b,f}^k}^{\tilde{A}\prec\tilde{B}\prec\tilde{F}} = E_{a,b,f}^{\tilde{A}\prec\tilde{B}\prec\tilde{F}},
\end{align}
so that our choice of process and of instruments above indeed allowed us to generate the causally ordered D-POVM $\mathbb{E}^{\tilde{A}\prec\tilde{B}\prec\tilde{F}} = (E_{a,b,f}^{\tilde{A}\prec\tilde{B}\prec\tilde{F}})_{a,b,f}$.

The last step in the argument, that takes the constructions for two causally ordered D-POVMs $\mathbb{E}^{\tilde{A}\prec\tilde{B}\prec\tilde{F}}$ and $\mathbb{E}^{\tilde{B}\prec\tilde{A}\prec\tilde{F}}$ to a convex mixture of those, is then similar to that in the bipartite case.

\section{Causally nonseparable D-POVMs from noncausal process matrix correlations}
\label{app:noncausal_imply_nonsepDPOVM}

If a process matrix $W^{AB}$ is noncausal, in the sense that it violates a causal inequality~\cite{oreshkov12} (or even if it is not ``extensibly causal'', i.e., if it violates a causal inequality after attaching some auxillary entangled state to it~\cite{oreshkov16}%
\footnote{For a non-extensibly-causal process matrix, the argument below applies in the same way, after attaching to $W^{AB}$ the auxillary entangled state that allows for the causal inequality violation.}%
), its causal nonseparability can be certified in a DI manner. Clearly, it can then also be certified in a semi-DI with trusted quantum inputs (SDI-QI) manner. Let us show how one can explicitly build a causally nonseparable D-POVM in such a situation.

Suppose that a causal inequality violation can be obtained from $W^{AB}$ by using some instruments $(M_{a|x}^A)_a$ and $(M_{b|y}^B)_b$, labelled by the classical inputs $x$ and $y$---i.e., that the correlations $P(a,b|x,y) = \Tr[(M_{a|x}^A\otimes M_{b|y}^B)^T W^{AB}]$ are noncausal~\cite{oreshkov12}. Let us then introduce some ancillary spaces $\HS^{\tilde{A}}$ and $\HS^{\tilde{B}}$ with (orthonormal) computational bases $\{\ket{x}^{\tilde{A}}\}_x$ and $\{\ket{y}^{\tilde{B}}\}_y$, and define
\begin{align}
M_a^{\tilde{A}A} = \sum_x \ketbra{x}{x}^{\tilde{A}}\otimes M_{a|x}^A, \notag \\
M_b^{\tilde{B}B} = \sum_y \ketbra{y}{y}^{\tilde{B}}\otimes M_{b|y}^B.
\end{align}
It is easily verified that $(M_a^{\tilde{A}A})_a$ and $(M_b^{\tilde{B}B})_b$ are valid quantum instruments (e.g., using $\Tr_{A_O} \sum_a M_{a|x}^A = \id^{A_I}$ for all $x$, one gets $\Tr_{A_O} \sum_a M_a^{\tilde{A}A} = \sum_x \ketbra{x}{x}^{\tilde{A}}\otimes\id^{A_I} = \id^{\tilde{A}A_I}$ as required).
With this choice, the induced D-POVM elements, obtained from Eq.~\eqref{eq:Eab}, are
\begin{align}
E_{a,b}^{\tilde{A}\tilde{B}}&=\sum_{x,y} ( \ketbra{x}{x}^{\tilde{A}}\otimes M_{a|x}^A \otimes \ketbra{y}{y}^{\tilde{B}}\otimes M_{b|y}^B ) * W^{AB} \notag \\
&=\sum_{x,y}P(a,b|x,y)\, \ketbra{x}{x}^{\tilde{A}}\otimes\ketbra{y}{y}^{\tilde{B}},
\end{align}
so that $P(a,b|x,y) = E_{a,b}^{\tilde{A}\tilde{B}} * (\ketbra{x}{x}^{\tilde{A}}\otimes\ketbra{y}{y}^{\tilde{B}})$.

We will see that the D-POVM $(E_{a,b}^{\tilde{A}\tilde{B}})_{a,b}$ thus obtained is causally nonseparable. For that, let us assume, by contradiction, that it is causally separable; its elements can then be decomposed as
\begin{align}
E_{a,b}^{\tilde{A}\tilde{B}} = q \, E_{a,b}^{\tilde{A}\prec \tilde{B}} + (1{-}q)\, E_{a,b}^{\tilde{B}\prec \tilde{A}}
\end{align}
for all $a,b$, with $q\in[0,1]$ and where $(E_{a,b}^{\tilde{A}\prec \tilde{B}})_{a,b}$ and $(E_{a,b}^{\tilde{B}\prec \tilde{A}})_{a,b}$ are D-POVMs compatible with $\tilde{A}\prec \tilde{B}$ and $\tilde{B}\prec \tilde{A}$, resp.
Then
\begin{align}
P(a,b|x,y) & = E_{a,b}^{\tilde{A}\tilde{B}} * (\ketbra{x}{x}^{\tilde{A}}\otimes\ketbra{y}{y}^{\tilde{B}}) \notag \\
& = q \,E_{a,b}^{\tilde{A}\prec \tilde{B}} * (\ketbra{x}{x}^{\tilde{A}}\otimes\ketbra{y}{y}^{\tilde{B}})  \notag \\
& \quad + (1{-}q)\, E_{a,b}^{\tilde{B}\prec \tilde{A}} * (\ketbra{x}{x}^{\tilde{A}}\otimes\ketbra{y}{y}^{\tilde{B}}) \notag \\
& = q\, p^{A\prec B}(a,b|x,y) + (1{-}q)\, p^{B\prec A} (a,b|x,y)
\end{align}
with $p^{A\prec B}(a,b|x,y) = E_{a,b}^{\tilde{A}\prec \tilde{B}} * (\ketbra{x}{x}^{\tilde{A}}\otimes\ketbra{y}{y}^{\tilde{B}})$ and $p^{B\prec A} (a,b|x,y) = E_{a,b}^{\tilde{B}\prec \tilde{A}} * (\ketbra{x}{x}^{\tilde{A}}\otimes\ketbra{y}{y}^{\tilde{B}})$, which are clearly (by definition of $E_{a,b}^{\tilde{A}\prec \tilde{B}}$ and $E_{a,b}^{\tilde{B}\prec \tilde{A}}$) causally ordered correlations. This contradicts the assumption that the correlations $P(a,b|x,y)$ are noncausal, and therefore indeed shows that the D-POVM $(E_{a,b}^{\tilde{A}\tilde{B}})_{a,b}$ constructed here is causally nonseparable.

\medskip

We note that the construction above generalises in a straightforward manner to the (2+$F$)-partite case (where causal correlations are those that can be written as a convex mixture of correlations compatible with the orders $A\prec B\prec F$ and $B\prec A\prec F$~\cite{oreshkov16,abbott16}).

\section{MDCI certification of causal nonseparability}
\label{app:MDCI_cert}

In the main text we considered imposing the form of Eq.~\eqref{eq:cstr_instr_v0} to Alice and Bob's instruments. We show here that trusting such a structure indeed allows one to certify the causal nonseparability of all bipartite causally nonseparable process matrices, and of all (2+$F$)-partite ``TTU-noncausal'' process matrices~\cite{bavaresco19}, in a measurement-device-and-channel-independent (MDCI) manner.

For this let us first note that the structure of Eq.~\eqref{eq:cstr_instr_v0} implies in particular that
\begin{align}
\Tr_{A_O} \!M_a^{\tilde{A}A} & = M_a^{\tilde{A}_I A_I} \!\otimes\! \id^{\tilde{A}_O}\!, \ \
\Tr_{B_O} \!M_b^{\tilde{B}B} = M_b^{\tilde{B}_I B_I} \!\otimes\! \id^{\tilde{B}_O} \label{eq:cstr_instr}
\end{align}
for all $a,b$, with $\sum_a M_a^{\tilde{A}_I A_I} = \id^{\tilde{A}_I A_I}$ and $\sum_b M_b^{\tilde{B}_I B_I} = \id^{\tilde{B}_I B_I}$.
These constraints, for the particular values of $a,b$ under consideration, are in fact sufficient to prove all the results below; hence, all references to Eq.~\eqref{eq:cstr_instr_v0} in the paper could in fact be replaced by the (weaker, but less directly motivated) constraints~\eqref{eq:cstr_instr} above.

\subsection{Bipartite case: MDCI certification of all causally nonseparable process matrices}

\subsubsection{Constraints on each D-POVM element \texorpdfstring{$E_{a,b}^{\tilde{A}\tilde{B}}$}{$E_{a,b}^{AB}$}}
\label{app:subsubsec:Cstr_each_Eab}

As we shall see, the constraints of Eq.~\eqref{eq:cstr_instr} imply that each D-POVM element $E_{a,b}^{\tilde{A}\tilde{B}}$ generated by a causally separable process matrix has a nontrivial structure.

To show this, let us start by considering a process matrix $W^{A\prec B} = W^{A\prec B_I}\otimes \id^{B_O}$ compatible with the order $A\prec B$.
Using Eq.~\eqref{eq:cstr_instr}, one then has
\begin{align}
E_{a,b}^{\tilde{A}\prec\tilde{B}} & = (M_a^{\tilde{A}A}\otimes M_b^{\tilde{B}B})*(W^{A\prec B_I}\otimes \id^{B_O}) \notag \\
& = (M_a^{\tilde{A}A}\otimes \Tr_{B_O} M_b^{\tilde{B}B})* W^{A\prec B_I} \notag \\
& = (M_a^{\tilde{A}A}\otimes M_b^{\tilde{B}_I B_I} \otimes \id^{\tilde{B}_O})* W^{A\prec B_I} \notag \\
& = E_{a,b}^{\tilde{A}\prec\tilde{B}_I} \otimes \id^{\tilde{B}_O} \label{eq:E00_AB_cstr}
\end{align}
with $E_{a,b}^{\tilde{A}\prec\tilde{B}_I} = (M_a^{\tilde{A}A}\otimes M_b^{\tilde{B}_I B_I})* W^{A\prec B_I} \ge 0$.
Similarly for a process matrix $W^{B\prec A} = W^{B\prec A_I}\otimes \id^{A_O}$ compatible with $B\prec A$, one gets $E_{a,b}^{\tilde{B}\prec\tilde{A}} = E_{a,b}^{\tilde{B}\prec\tilde{A}_I} \otimes \id^{\tilde{A}_O}$ with $E_{a,b}^{\tilde{B}\prec\tilde{A}_I} = (M_a^{\tilde{A}_IA_I}\otimes M_b^{\tilde{B} B})* W^{B\prec A_I} \ge 0$.
More generally, starting with a causally separable process matrix $W^{AB}$ as in Eq.~\eqref{eq:csep2} of the main text, one finds that $E_{a,b}^{\tilde{A}\tilde{B}}$ necessarily decomposes as
\begin{align}
E_{a,b}^{\tilde{A}\tilde{B}} & = q\,E_{a,b}^{\tilde{A}\prec\tilde{B}_I} \otimes \id^{\tilde{B}_O} + (1{-}q)\,E_{a,b}^{\tilde{B}\prec\tilde{A}_I} \otimes \id^{\tilde{A}_O} \label{eq:csep_POVM_00_app}
\end{align}
for some $E_{a,b}^{\tilde{A}\prec\tilde{B}_I}, E_{a,b}^{\tilde{B}\prec\tilde{A}_I} \ge 0$, as in Eq.~\eqref{eq:csep_POVM_00}. \emph{A contrario}, if $E_{a,b}^{\tilde{A}\tilde{B}}$ cannot be decomposed in such a way, then one can conclude that $W^{AB}$ is causally nonseparable.

\medskip

Let us note that imposing some specific structure to (both) Alice and Bob's instruments, as in Eq.~\eqref{eq:cstr_instr_v0} or~\eqref{eq:cstr_instr}, was required to reach this conclusion: without such assumptions, the individual D-POVM elements induced by causally separable process matrices do not necessarily decompose as in Eq.~\eqref{eq:csep_POVM_00_app} above.%
\footnote{Indeed, without requiring Eq.~\eqref{eq:cstr_instr}, up to normalisation any given PSD matrix $\tilde{E} \in \L(\HS^{\tilde{A}\tilde{B}})$ can be obtained from a causally separable (in fact, even a nonsignaling) process matrix $W^{AB}$ as $\tilde{E} \propto E_{0,0}^{\tilde{A}\tilde{B}} = (M_0^{\tilde{A}A}\otimes M_0^{\tilde{B}B})*W^{AB}$, for some choice of CP trace-non-increasing (TNI) maps $M_0^{\tilde{A}A}$ and $M_0^{\tilde{B}B}$ (corresponding to the specific outputs $a=b=0$ of some instruments $(M_a^{\tilde{A}A})_a$ and $(M_b^{\tilde{B}B})_b$). 
\\
To see this, take for instance $\HS^{A_I}$ and $\HS^{B_I}$ to be isomorphic to $\HS^{\tilde{A}} = \HS^{\tilde{A}_I\tilde{A}_O}$ and $\HS^{\tilde{B}} = \HS^{\tilde{B}_I\tilde{B}_O}$, resp., take some trivial (one-dimensional) $\HS^{A_O}$ and $\HS^{B_O}$, take $M_0^{\tilde{A}A}$ and $M_0^{\tilde{B}B}$ to be projections onto the maximally entangled states $\ket{\Phi^+}^{\tilde{A}A_I} = \frac{1}{\sqrt{d_{\tilde{A}}}}\dket{\id}^{\tilde{A}A_I}$ and $\ket{\Phi^+}^{\tilde{B}B_I} = \frac{1}{\sqrt{d_{\tilde{B}}}}\dket{\id}^{\tilde{B}B_I}$, resp.\ (which indeed fail to satisfy Eq.~\eqref{eq:cstr_instr} as soon as $\HS^{\tilde{A}_O}$ or $\HS^{\tilde{B}_O}$ are nontrivial), and take $W^{AB} = \tilde{\rho}^{A_IB_I} \in \L(\HS^{A_IB_I})$ (a quantum state) to be the same as $\tilde{E} / \Tr(\tilde{E})$, but written in the spaces $\HS^{A_I}, \HS^{B_I}$ (more formally: $\tilde{\rho}^{A_IB_I} = (\dketbra{\id}{\id}^{\tilde{A}A_I} \otimes \dketbra{\id}{\id}^{\tilde{B}B_I}) * \tilde{E} / \Tr(\tilde{E})$). Then Eq.~\eqref{eq:Eab} indeed gives (up to normalisation) the desired matrix $E_{0,0}^{\tilde{A}\tilde{B}} =  \tilde{E} / [d_{\tilde{A}}d_{\tilde{B}}\Tr(\tilde{E})]$.
\\
If the structure of Eq.~\eqref{eq:cstr_instr} is imposed to one party only (say Alice) then again any PSD matrix $\tilde{E} \in \L(\HS^{\tilde{A}\tilde{B}})$ can be obtained, up to normalisation, as $\tilde{E} \propto E_{0,0}^{\tilde{A}\tilde{B}} = (M_0^{\tilde{A}A}\otimes M_0^{\tilde{B}B})*W^{AB}$, now for some $W^{AB}$ with a fixed order (say $A\prec B$). To see this, take now $\HS^{A_I}$, $\HS^{A_O}$ and $\HS^{B_I^{(0)}}$ isomorphic to $\HS^{\tilde{A}_I}$, $\HS^{\tilde{A}_O}$ and $\HS^{\tilde{B}}$, resp., introduce some spaces $\HS^{B_I^{(1)}}, \HS^{B_I^{(2)}}$ both isomorphic to $\HS^{\tilde{A}_O}$, define $\HS^{B_I} = \HS^{B_I^{(0)}B_I^{(1)}B_I^{(2)}}$ and take some trivial $\HS^{B_O}$; then take $W^{AB} = W^{A\prec B_I} \propto \tilde{\rho}^{A_IB_I^{(1)}B_I^{(0)}} \otimes \dketbra{\id}{\id}^{A_OB_I^{(2)}}$ with $\tilde{\rho}^{A_IB_I^{(1)}B_I^{(0)}} \in \L(\HS^{A_IB_I^{(1)}B_I^{(0)}})$ the same as $\tilde{E} \in \L(\HS^{\tilde{A}_I\tilde{A}_O\tilde{B}})$, but written in the spaces $\HS^{A_IB_I^{(1)}B_I^{(0)}}$; and take $M_0^{\tilde{A}A} = \ketbra{\Phi^+}{\Phi^+}^{\tilde{A}_IA_I} \otimes \dketbra{\id}{\id}^{\tilde{A}_OA_O}$ (that indeed satisfies Eq.~\eqref{eq:cstr_instr}) and $M_0^{\tilde{B} B_I} = \ketbra{\Phi^+}{\Phi^+}^{\tilde{B}B_I^{(0)}} \otimes \ketbra{\Phi^+}{\Phi^+}^{B_I^{(1)}B_I^{(2)}}$ (that fails to satisfy Eq.~\eqref{eq:cstr_instr} if $\HS^{\tilde{B}_O}$ is nontrivial).
\label{footnote:anyE00}}

One may also wonder if the operators $E_{a,b}^{\tilde{A}\prec\tilde{B}_I}$ and $E_{a,b}^{\tilde{B}\prec\tilde{A}_I}$ obtained above have any further structure---e.g., given how $\tilde{A}$ decomposes here into $\tilde{A}_I \tilde{A}_O$, whether $E_{a,b}^{\tilde{A}\prec\tilde{B}_I}$ should also satisfy $_{[1-\tilde{A}_O]\tilde{B}_I}E_{a,b}^{\tilde{A}\prec\tilde{B}_I}=0$ (just like $W^{A\prec B_I}$ satisfies $_{[1-A_O]B_I}W^{A\prec B_I}=0$). This is however not the case in general.%
\footnote{Indeed, any given PSD matrix $\tilde{E} \in \L(\HS^{\tilde{A}_I\tilde{A}_O\tilde{B}_I})$ can be obtained, up to normalisation, as $\tilde{E} \propto E_{0,0}^{\tilde{A}\prec\tilde{B}_I} = (M_0^{\tilde{A}A}\otimes M_0^{\tilde{B}_I B_I})* W^{A\prec B_I}$ from a causally ordered $W^{A\prec B_I}$ and a specific choice of CP TNI maps $M_0^{\tilde{A}A}, M_0^{\tilde{B}_I B_I}$ satisfying Eq.~\eqref{eq:cstr_instr}.
This can be seen using the same construction as in the last paragraph in Footnote~\ref{footnote:anyE00}, after replacing $\tilde{B}$ by just $\tilde{B}_I$ (as we have here a trivial $\HS^{\tilde{B}_O}$). \label{footnote:anyE00ABI}}

\subsubsection{Certifying any bipartite causally nonseparable process matrix}

We now show that any bipartite causally nonseparable process matrix $W^{AB}$ can generate a D-POVM that is \emph{not} of the form of Eq.~\eqref{eq:csep_POVM_00_app}, using instruments that decompose as in Eq.~\eqref{eq:cstr_instr_v0}.

To this end, let us take ancillary quantum input spaces $\HS^{\tilde{A}_I}, \HS^{\tilde{A}_O}, \HS^{\tilde{B}_I}$ and $\HS^{\tilde{B}_O}$ that are isomorphic to $\HS^{A_I}, \HS^{A_O}, \HS^{B_I}$ and $\HS^{B_O}$, resp., and consider the instrument elements (that, for $a=b=0$, indeed satisfy Eq.~\eqref{eq:cstr_instr} and the stronger Eq.~\eqref{eq:cstr_instr_v0})
\begin{align}
M_0^{\tilde{A}A} & = {\textstyle \frac{1}{d_{A_I}}}\dketbra{\id}{\id}^{\tilde{A}_IA_I} \otimes \dketbra{\id}{\id}^{\tilde{A}_OA_O}, \notag \\
M_0^{\tilde{B}B} & = {\textstyle \frac{1}{d_{B_I}}}\dketbra{\id}{\id}^{\tilde{B}_IB_I} \otimes \dketbra{\id}{\id}^{\tilde{B}_OB_O}, \label{eq:id_instruments}
\end{align}
where $d_{A_I}, d_{B_I}$ are the dimensions of $\HS^{A_I}$ and $\HS^{B_I}$, resp.
These instruments correspond (e.g., for Alice, and analogously for Bob) to a projection onto the (normalised) maximally entangled state $\frac{1}{\sqrt{d_{A_I}}}\dket{\id}^{\tilde{A}_IA_I}$ and an identity channel from $\HS^{\tilde{A}_O}$ to $\HS^{A_O}$.
As the projection onto $\frac{1}{\sqrt{d_{A_I}}}\dket{\id}^{\tilde{A}_IA_I}$ also essentially amounts to a (post-selected) identity channel from $\HS^{\tilde{A}_I}$ to $\HS^{A_I}$, these operations effectively ``teleport'' $W^{AB}$ onto to ancillary spaces so that the induced D-POVM element $E_{0,0}^{\tilde{A}\tilde{B}}$ is found to be formally the same, up to a normalisation factor $\frac{1}{d_{A_I}d_{B_I}}$, as the process matrix $W^{AB}$, but written in the spaces $\HS^{\tilde{A}_I\tilde{A}_O}, \HS^{\tilde{B}_I\tilde{B}_O}$.

Suppose now that $W^{AB}$ is causally nonseparable, so that it \emph{cannot} be decomposed as in Eq.~\eqref{eq:csep2}. Now, recall that such a decomposition remains impossible even if we do not require $W^{A\prec B_I}, W^{B\prec A_I} \ge 0$ to be valid process matrices \emph{a priori}: see the discussion around Eqs.~\eqref{eq:csep2_app}--\eqref{eq:constr_W_AprecBI} in Sec.~\ref{app:validity_cstrs_W}.%
\footnote{This subtlety in the argument prevents our results from generalising in a straightforward manner to the multipartite case; see Sec.~\ref{app:validity_cstrs_W}. \label{footnote:subtlety}}
Translating this onto $E_{0,0}^{\tilde{A}\tilde{B}}$, this implies that the latter cannot be decomposed as in Eq.~\eqref{eq:csep_POVM_00_app}, for any $E_{0,0}^{\tilde{A}\prec\tilde{B}_I}, E_{0,0}^{\tilde{B}\prec\tilde{A}_I} \ge 0$, as claimed above.

\medskip

As discussed in the previous subsection, verifying that $E_{0,0}^{\tilde{A}\tilde{B}}$ is not of the form of Eq.~\eqref{eq:csep_POVM_00_app} implies that $W^{AB}$ is causally nonseparable. This verification can be done with similar techniques to the use of ``witnesses of causal nonseparability''~\cite{araujo15,branciard16a}, see Sec.~\ref{app:cones_charact}. Such witnesses can be measured in practice by using complete enough sets of quantum input sets $\{\rho_x^{\tilde{A}}\}_x$ and $\{\rho_y^{\tilde{B}}\}_y$, as discussed in the main text. Because we need here to trust that Alice and Bob's instruments are of the form of Eq.~\eqref{eq:cstr_instr_v0}, this certification is MDCI.

\subsection{(2+$F$)-partite case: MDCI certification of all TTU-noncausal processes}
\label{app:MDCI_2F_TTU}

\subsubsection{Constraints on the D-POVM elements  \texorpdfstring{$(E_{a,b,f}^{\tilde{A}\tilde{B}\tilde{F}})_f$}{$(E_{a,b,f}^{ABF})_f$} for fixed $a,b$, and all $f$}
\label{app:MDCI_2F_TTU_allQIs}

Contrarily to the bipartite case, in the (2+$F$)-partite case, the requirement that Alice and Bob's instruments are of the form of Eq.~\eqref{eq:cstr_instr_v0} (or even, that they satisfy Eq.~\eqref{eq:cstr_instr} for some fixed $a,b$)%
\footnote{Note that since Fiona has no output Hilbert space, the analogous condition to Eq.~\eqref{eq:cstr_instr_v0} for her instrument is automatically satisfied, for some trivial $\tilde{F}_O$.}
is not enough to impose any specific structure to each individual D-POVM element $E_{a,b,f}^{\tilde{A}\tilde{B}\tilde{F}}$.%
\footnote{Indeed, any given PSD matrix $\tilde{E} \in \L(\HS^{\tilde{A}_I\tilde{A}_O\tilde{B}_I\tilde{B}_O\tilde{F}})$ can be obtained, up to normalisation, from a process matrix $W^{ABF}$ compatible with both causal orders $A\prec B\prec F$ and $B\prec A\prec F$, for some choice of CP TNI maps $M_0^{\tilde{A}A}$, $M_0^{\tilde{B}B}$ satisfying Eq.~\eqref{eq:cstr_instr_v0} and some $M_0^{\tilde{F}F}$, as $\tilde{E} \propto E_{0,0,0}^{\tilde{A}\tilde{B}\tilde{F}} = (M_0^{\tilde{A}A}\otimes M_0^{\tilde{B}B}\otimes M_0^{\tilde{F}F})*W^{ABF}$. An explicit construction similar in spirit to that referred to in Footnote~\ref{footnote:anyE00ABI} could indeed be provided here.}
Nevertheless, some nontrivial structure is recovered when considering all of Fiona's outcomes $f$.

To see this, consider first a process matrix $W^{A\prec B\prec F}$ compatible with the order $A\prec B\prec F$ (hence satisfying in particular $\Tr_{F} W^{A\prec B\prec F} = W^{A\prec B_I}\otimes \id^{B_O}$ for some process matrix $W^{A\prec B_I}$). In a similar way to the bipartite case, see Eq.~\eqref{eq:E00_AB_cstr}, summing over $f$ (for some fixed $a,b$, and using the fact that $\sum_f M_f^{\tilde{F}F} = \id^{\tilde{F}F}$) one gets:
\begin{align}
\sum_f E_{a,b,f}^{\tilde{A}\prec\tilde{B}\prec\tilde{F}} & = (M_a^{\tilde{A}A}\otimes M_b^{\tilde{B}B}\otimes\sum_f M_f^{\tilde{F}F}) * W^{A\prec B\prec F} \notag \\
& = (M_a^{\tilde{A}A}\otimes M_b^{\tilde{B}B}) * \Tr_{F} W^{A\prec B\prec F} \otimes \id^{\tilde{F}} \notag \\
& = (M_a^{\tilde{A}A}\otimes M_b^{\tilde{B}B}) * (W^{A\prec B_I}\otimes \id^{B_O}) \otimes \id^{\tilde{F}} \notag \\
& = E_{a,b}^{\tilde{A}\prec\tilde{B}_I} \otimes \id^{\tilde{B}_O\tilde{F}} \label{eq:sum_f_Eabf}
\end{align}
with $E_{a,b}^{\tilde{A}\prec\tilde{B}_I} = (M_a^{\tilde{A}A}\otimes M_b^{\tilde{B}_I B_I})* W^{A\prec B_I} \ge 0$.
Similarly for a process matrix $W^{B\prec A\prec F}$ compatible with $B\prec A\prec F$, one gets
\begin{align}
\sum_f E_{a,b,f}^{\tilde{B}\prec\tilde{A}\prec\tilde{F}} & = E_{a,b}^{\tilde{B}\prec\tilde{A}_I} \otimes \id^{\tilde{A}_O\tilde{F}}
\end{align}
with $E_{a,b}^{\tilde{B}\prec\tilde{A}_I} = (M_a^{\tilde{A}_IA_I}\otimes M_b^{\tilde{B} B})* W^{B\prec A_I} \ge 0$.

More generally, starting with a causally separable process matrix $W^{ABF} = q\, W^{A\prec B\prec F} + (1{-}q)\, W^{B\prec A\prec F}$, one finds that the $E_{a,b,f}^{\tilde{A}\tilde{B}\tilde{F}}$'s (for some fixed $a,b$) necessarily decompose as
\begin{align}
E_{a,b,f}^{\tilde{A}\tilde{B}\tilde{F}} & = q\,E_{a,b,f}^{\tilde{A}\prec\tilde{B}\prec\tilde{F}} + (1{-}q)\,E_{a,b,f}^{\tilde{B}\prec\tilde{A}\prec\tilde{F}} \quad \forall\,f \label{eq:cstr_E00f_csep_1}
\end{align}
for some $E_{a,b,f}^{\tilde{A}\prec\tilde{B}\prec\tilde{F}}, E_{a,b,f}^{\tilde{B}\prec\tilde{A}\prec\tilde{F}} \ge 0$ satisfying
\begin{align}
\sum_f E_{a,b,f}^{\tilde{A}\prec\tilde{B}\prec\tilde{F}} & = E_{a,b}^{\tilde{A}\prec\tilde{B}_I} \otimes \id^{\tilde{B}_O\tilde{F}}, \notag \\
\sum_f E_{a,b,f}^{\tilde{B}\prec\tilde{A}\prec\tilde{F}} & = E_{a,b}^{\tilde{B}\prec\tilde{A}_I} \otimes \id^{\tilde{A}_O\tilde{F}} \label{eq:cstr_E00f_csep_2}
\end{align}
for some $E_{a,b}^{\tilde{A}\prec\tilde{B}_I}, E_{a,b}^{\tilde{B}\prec\tilde{A}_I} (\ge 0)$.

\medskip

As in the bipartite case above, we note that imposing some specific structure to Alice and Bob's instruments, as in Eq.~\eqref{eq:cstr_instr_v0} or~\eqref{eq:cstr_instr}, was required to reach this conclusion,%
\footnote{Without requiring Eq.~\eqref{eq:cstr_instr}, up to normalisation any family of PSD matrices $(\tilde{E}_f)_f$ in $\L(\HS^{\tilde{A}\tilde{B}})$ can be obtained from a nonsignaling process matrix (a quantum state) $W^{ABF}$,  as $\tilde{E}_f \propto E_{0,0,f}^{\tilde{A}\tilde{B}\tilde{F}} = (M_0^{\tilde{A}A}\otimes M_0^{\tilde{B}B}\otimes M_f^{\tilde{F}F})*W^{ABF}$: take, e.g., $\HS^{A_I}$, $\HS^{B_I}$, $\HS^{A_O}$, $\HS^{B_O}$, $M_0^{\tilde{A}A}$ and $M_0^{\tilde{B}B}$ as in Footnote~\ref{footnote:anyE00}, $\HS^F = \HS^{F^{(1)}F^{(2)}}$ with $\HS^{F^{(1)}}$ isomorphic to $\HS^{\tilde{F}}$ and $\HS^{F^{(2)}}$ a Hilbert space with an orthonormal basis $\{\ket{f}^{F^{(2)}}\}_f$, $M_f^{\tilde{F}F} \propto \dketbra{\id}{\id}^{\tilde{F}F^{(1)}}\otimes\ketbra{f}{f}^{F^{(2)}}$ and $W^{ABF} \propto \sum_f \tilde{\rho}_f^{A_IB_IF^{(1)}}\otimes\ketbra{f}{f}^{F^{(2)}} \in \L(\HS^{A_IB_IF})$ with each $\tilde{\rho}_f^{A_IB_IF^{(1)}}$ being (up to normalisation) the same as $\tilde{E}_f$, but written in the spaces $\HS^{A_I}, \HS^{B_I}, \HS^{F^{(1)}aa}$.}
and that the operators $E_{a,b}^{\tilde{A}\prec\tilde{B}_I}$ and $E_{a,b}^{\tilde{B}\prec\tilde{A}_I}$ above have no further specific structure in general.

\subsubsection{With classical inputs for Fiona: constraints on the D-POVM elements  \texorpdfstring{$(E_{a,b,f|z}^{\tilde{A}\tilde{B}})_{f,z}$}{$(E_{a,b,f|z}^{AB})_{f,z}$} for fixed $a,b$, and all $f,z$}

It remains an open problem to characterise precisely which causally nonseparable (2+$F$)-partite process matrices can generate families of D-POVM elements $(E_{a,b,f}^{\tilde{A}\tilde{B}\tilde{F}})_f$ (for some fixed $a,b$) that \emph{cannot} be decomposed as in Eqs.~\eqref{eq:cstr_E00f_csep_1}--\eqref{eq:cstr_E00f_csep_2}. As a generalisation to our result in the bipartite case, we find that this is at least the case for all process matrices that are said to be ``TTU-noncausal''~\cite{bavaresco19}.

Here ``TTU'' stands for ``Trusted-Trusted-Untrusted'', and refers to a (2+$F$)-partite scenario where Alice and Bob's instruments are trusted, while Fiona's measurement is not: her choice of measurement is simply labelled by a classical variable $z$. TTU-noncausal process matrices are those whose causal nonseparability can be certified in such a TTU manner (see next subsection). To establish our result below, we will correspondingly consider a situation where Fiona has some classical, rather than quantum inputs.

As before, we start by looking at what the causal separability of a process matrix implies on the structure of the induced D-POVMs. 
Building on the previous calculations, this can equivalently be done either by considering that Fiona's possible inputs are orthogonal states $\ket{z}^{\tilde{F}}$ and that her POVM elements are of the form $M_f^{F\tilde{F}} = \sum_z M_{f|z}^F \otimes\ketbra{z}{z}^{\tilde{F}}$ for some POVMs $\{M_{f|z}^F\}_f$ conditioned by $z$; or by removing Fiona's quantum inputs and conditioning ``by hand'' all previous calculations by Fiona's classical input, noting that Alice and Bob's marginal D-POVMs should not depend on Fiona's input $z$, which comes in their causal future. Eqs.~\eqref{eq:cstr_E00f_csep_1}--\eqref{eq:cstr_E00f_csep_2} thus provide the following conditions, for all subsets of D-POVM elements $(E_{a,b,f|z}^{\tilde{A}\tilde{B}} = E_{a,b,f}^{\tilde{A}\tilde{B}\tilde{F}} * \ketbra{z}{z}^{\tilde{F}} = (M_a^{\tilde{A}A}\otimes M_b^{\tilde{B}B}\otimes M_{f|z}^F)*W^{ABF})_{f,z}$ (for fixed $a,b$) induced by a causally separable process matrix:
\begin{align}
E_{a,b,f|z}^{\tilde{A}\tilde{B}} & = q\,E_{a,b,f|z}^{\tilde{A}\prec\tilde{B}} + (1{-}q)\,E_{a,b,f|z}^{\tilde{B}\prec\tilde{A}} \quad \forall\,f,z, \label{eq:cstr_E00fz_csep_1}
\end{align}
\begin{align}
\sum_f E_{a,b,f|z}^{\tilde{A}\prec\tilde{B}} & = E_{a,b}^{\tilde{A}\prec\tilde{B}_I} \otimes \id^{\tilde{B}_O} \quad \forall\,z, \notag \\
\sum_f E_{a,b,f|z}^{\tilde{B}\prec\tilde{A}} & = E_{a,b}^{\tilde{B}\prec\tilde{A}_I} \otimes \id^{\tilde{A}_O} \quad \forall\,z \label{eq:cstr_E00fz_csep_2}
\end{align}
for some $E_{a,b,f|z}^{\tilde{A}\prec\tilde{B}}, E_{a,b,f|z}^{\tilde{B}\prec\tilde{A}}, E_{a,b}^{\tilde{A}\prec\tilde{B}_I}, E_{a,b}^{\tilde{B}\prec\tilde{A}_I} \ge 0$.

If no such decomposition exists, then this implies that the process matrix under consideration is causally nonseparable. (Note that this also implies that the corresponding sets of D-POVM elements $(E_{a,b,f}^{\tilde{A}\tilde{B}\tilde{F}})_f$, that include the classical inputs, cannot be decomposed as in Eqs.~\eqref{eq:cstr_E00f_csep_1}--\eqref{eq:cstr_E00f_csep_2}.)

\subsubsection{Certifying any (2+$F$)-partite TTU-noncausal process matrix}
\label{app:anyTTU}

As defined in Ref.~\cite{bavaresco19}, a ``TTU-assemblage'' is a set of PSD matrices $(w_{f|z}^{AB})_{f,z}$, with each $w_{f|z}^{AB} \in \L(\HS^{AB})$, such that
\begin{align}
\sum_f w_{f|z}^{AB} = W^{AB} \qquad \forall\,z, \label{eq:def_TTU_assemblage}
\end{align}
for some bipartite process matrix $W^{AB}$.
A TTU-assemblage is said to be ``causal'' if it can be decomposed as a convex mixture (for some $q \in [0,1]$)
\begin{align}
w_{f|z}^{AB} = q\, w_{f|z}^{A\prec B} + (1{-}q)\, w_{f|z}^{B\prec A} \qquad \forall\,f,z, \label{eq:causalTTU_decomp}
\end{align}
in terms of matrices PSD $w_{f|z}^{A\prec B} , w_{f|z}^{B\prec A} \ge 0$ satisfying
\begin{align}
\sum_f w_{f|z}^{A\prec B} & = W^{A\prec B_I} \otimes \id^{B_O} \qquad \forall\,z, \notag \\
\sum_f w_{f|z}^{B\prec A} & = W^{B\prec A_I} \otimes \id^{A_O} \qquad \forall\,z, \label{eq:causalTTU_decomp_cstr}
\end{align}
for some causally ordered (valid) process matrices $W^{A\prec B_I}$ and $W^{B\prec A_I}$.

If no such decomposition exists, then the TTU-assemblage is noncausal.
As in the bipartite case (cf.\ the discussion at the end of Sec.~\ref{app:validity_cstrs_W} above), we note that a noncausal TTU-assemblage can also not be decomposed as in Eqs.~\eqref{eq:causalTTU_decomp} and~\eqref{eq:causalTTU_decomp_cstr} above, even if we do not assume \emph{a priori} that $W^{A\prec B_I}$ and $W^{B\prec A_I}$ are valid process matrices (as their validity condition would anyway be implied by Eqs.~\eqref{eq:def_TTU_assemblage}--\eqref{eq:causalTTU_decomp_cstr}, as in Footnote~\ref{footnote:valid_a_fortiori}).

\medskip

A TTU-assemblage $(w_{f|z}^{AB})_{f,z}$ is typically obtained as a ``process TTU-assemblage'', that is, starting from a (2+$F$)-partite process matrix $W^{ABF}$, by letting Fiona apply some POVMs $(M_{f|z}^F)_f$ (for some classical inputs $z$) and defining the matrices $w_{f|z}^{AB} = M_{f|z}^F * W^{ABF}$. (Such a set $(w_{f|z}^{AB})_{f,z}$ indeed satisfies Eq.~\eqref{eq:def_TTU_assemblage}, with $W^{AB} = \Tr_F W^{ABF}$.) A process matrix that can generate a noncausal TTU-assemblage in such a way is said to be ``TTU-noncausal''~\cite{bavaresco19}. Note that only causally nonseparable process matrices can generate noncausal process TTU-assemblages, so that certifying TTU-noncausality implies a certification of causal nonseparability.

\medskip

Consider, as in the bipartite case, some quantum input spaces $\HS^{\tilde{A}_I}, \HS^{\tilde{A}_O}, \HS^{\tilde{B}_I}$ and $\HS^{\tilde{B}_O}$ that are isomorphic to $\HS^{A_I}, \HS^{A_O}, \HS^{B_I}$ and $\HS^{B_O}$, resp., and the CP maps $M_0^{\tilde{A}A}$ and $M_0^{\tilde{B}B}$ of Eq.~\eqref{eq:id_instruments} for Alice and Bob. As before, we find that the induced D-POVM elements $E_{0,0,f|z}^{\tilde{A}\tilde{B}} = (M_0^{\tilde{A}A}\otimes M_0^{\tilde{B}B}\otimes M_{f|z}^F)*W^{ABF} = (M_0^{\tilde{A}A}\otimes M_0^{\tilde{B}B})*w_{f|z}^{AB}$ are formally the same, up to a normalisation factor $\frac{1}{d_{A_I}d_{B_I}}$, as the matrices $w_{f|z}^{AB}$ of the TTU-assemblage generated by the process matrix $W^{ABF}$ and Fiona's POVMs $(M_{f|z}^F)_f$, but written in the spaces $\tilde{A}_I, \tilde{A}_O, \tilde{B}_I$ and $\tilde{B}_O$.

Suppose now that the TTU-assemblage $(w_{f|z}^{AB})_{f,z}$ is noncausal---i.e., that $W^{ABF}$ is TTU-noncausal---so that it cannot be decomposed as in Eqs.~\eqref{eq:causalTTU_decomp}--\eqref{eq:causalTTU_decomp_cstr} for any $w_{f|z}^{A\prec B}, w_{f|z}^{B\prec A}, W^{A\prec B_I}, W^{B\prec A_I} \ge 0$. Now, recall from the remark above that such a decomposition remains impossible even if we do not require $W^{A\prec B_I}, W^{B\prec A_I} \ge 0$ to be valid process matrices \emph{a priori}.
Translating this onto the $E_{0,0,f|z}^{\tilde{A}\tilde{B}}$'s, this implies that the latter cannot be decomposed as in Eqs.~\eqref{eq:cstr_E00fz_csep_1}--\eqref{eq:cstr_E00fz_csep_2}, for any $E_{0,0,f|z}^{\tilde{A}\prec\tilde{B}}, E_{0,0,f|z}^{\tilde{B}\prec\tilde{A}}, E_{0,0}^{\tilde{A}\prec\tilde{B}_I}, E_{0,0}^{\tilde{B}\prec\tilde{A}_I} \ge 0$.

\medskip

Hence, we have shown that any TTU-noncausal (2+$F$)-partite process matrix $W^{ABF}$ can generate some subsets of D-POVM elements $(E_{0,0,f|z}^{\tilde{A}\tilde{B}})_{f,z}$ that cannot be decomposed as in Eqs.~\eqref{eq:cstr_E00fz_csep_1}--\eqref{eq:cstr_E00fz_csep_2}. Verifying that such a decomposition is impossible can again be done with similar techniques to the use of ``witnesses of causal nonseparability''~\cite{araujo15,branciard16a} (see Sec.~\ref{app:cones_charact}), which can be measured in practice. Here we do not need to trust Fiona's POVMs but we need to trust that Alice and Bob's instruments are of the form of Eq.~\eqref{eq:cstr_instr_v0} of the main text, so that this certification of causal nonseparability is MDCI for Alice and Bob, but fully DI for Fiona (``MDCI-MDCI-DI'').

The quantum switch is an example of a TTU-noncausal process~\cite{bavaresco19}, whose TTU-noncausality---and hence, whose causal nonseparability---can thus be certified in such a way. We note that for the quantum switch Fiona can apply a single, fixed POVM $(M_f^F)_f$, with no classical input $z$; see also Sec.~\ref{app:QSwitch}.

\subsection{(2+$F$)-partite case: MDCI certification of all TUU-noncausal processes}
\label{app:MDCI_2F_TUU}

Let us now consider a ``TUU'' (for ``Trusted-Untrusted-Untrusted'') scenario~\cite{bavaresco19}, where Bob's instrument is also untrusted. Clearly, if a (2+$F$)-partite process matrix can be certified to be causally nonseparable in a TUU manner---in other words, if it is ``TUU-noncausal''---then it can also be certified in a TTU manner---i.e., it is also TTU-noncausal. From the previous section, we conclude that in a scenario with trusted quantum inputs it can also be certified in a MDCI manner for Alice and Bob, and fully DI for Fiona (``MDCI-MDCI-DI'').
Unsurprisingly this result can be strengthened: in the latter scenario Bob's devices can also be fully unstrusted, so that the certification can be MDCI for Alice only, and fully DI for both Bob and Fiona (``MDCI-DI-DI'').

\subsubsection{With classical inputs for Bob and Fiona: constraints on the D-POVM elements  \texorpdfstring{$(E_{a,b,f|y,z}^{\tilde{A}})_{b,f,y,z}$}{$(E_{a,b,f|y,z}^A)_{b,f,y,z}$} for fixed $a$, and all $b,f,y,z$}

Assume that Alice's instrument satisfies the constraint of Eq.~\eqref{eq:cstr_instr_v0} (or Eq.~\eqref{eq:cstr_instr} directly, for some particular $a$), and let us directly consider here the relevant case with classical inputs $y,z$ for Bob and Fiona. (The case where they still have quantum inputs could also be considered, in a similar manner to Sec.~\ref{app:MDCI_2F_TTU_allQIs} above.)

Considering a process matrix $W^{A\prec B\prec F}$ compatible with the order $A\prec B\prec F$ (such that $\Tr_{F} W^{A\prec B\prec F} = W^{A\prec B_I}\otimes \id^{B_O}$ and $\Tr_{B_I} W^{A\prec B_I} = \rho^{A_I} \otimes \id^{A_O}$), the induced D-POVM elements $E_{a,b,f|y,z}^{\tilde{A}\,[A\prec B]} = (M_a^{\tilde{A}A}\otimes M_{b|y}^B\otimes M_{f|z}^F) * W^{A\prec B\prec F}$ satisfy, for some fixed $a$ (similarly to Eq.~\eqref{eq:sum_f_Eabf} or~\eqref{eq:cstr_E00fz_csep_2}, using the TP condition $\Tr_{B_O}\sum_bM_{b|y}^B = \id^{B_I}$ and Eq.~\eqref{eq:cstr_instr})
\begin{align}
\sum_f E_{a,b,f|y,z}^{\tilde{A}\,[A\prec B]}
= E_{a,b|y}^{\tilde{A}\,[A\prec B]} \qquad & \forall\, b,y,z, \notag \\
\sum_b E_{a,b|y}^{\tilde{A}\,[A\prec B]} = E_a^{\tilde{A}_I\,[A\prec B]}\otimes\id^{\tilde{A}_O} \qquad & \forall\, y, \label{eq:cstr_Eabfyz_AprecB}
\end{align}
with $E_{a,b|y}^{\tilde{A}\,[A\prec B]} = (M_a^{\tilde{A}A}\otimes \Tr_{B_O}M_{b|y}^B)* W^{A\prec B_I}$ and $E_a^{\tilde{A}_I\,[A\prec B]} = M_a^{\tilde{A}_IA_I} * \rho^{A_I}$.

Considering a process matrix $W^{B\prec A\prec F}$ compatible with the order $B\prec A\prec F$ (such that $\Tr_{F} W^{B\prec A\prec F} = W^{B\prec A_I}\otimes \id^{A_O}$), the induced D-POVM elements $E_{a,b,f|y,z}^{\tilde{A}\,[B\prec A]} = (M_a^{\tilde{A}A}\otimes M_{b|y}^B\otimes M_{f|z}^F) * W^{B\prec A\prec F}$ satisfy, again for some fixed $a$,
\begin{align}
\sum_f E_{a,b,f|y,z}^{\tilde{A}\,[B\prec A]}
& = E_{a,b|y}^{\tilde{A}_I\,[B\prec A]}\otimes\id^{\tilde{A}_O} \qquad \forall\, b,y,z, \label{eq:cstr_Eabfyz_BprecA}
\end{align}
with $E_{a,b|y}^{\tilde{A}_I\,[B\prec A]} = (M_a^{\tilde{A}_IA_I}\otimes M_{b|y}^B)* W^{B\prec A_I}$.

\medskip

Hence, the D-POVM induced by a causally separable process matrix $W^{ABF} = q\,W^{A\prec B\prec F} + (1{-}q)\,W^{B\prec A\prec F}$ is such that its elements $E_{a,b,f|y,z}^{\tilde{A}}$ can be decomposed (for some fixed $a$) as
\begin{align}
E_{a,b,f|y,z}^{\tilde{A}} = q \, E_{a,b,f|y,z}^{\tilde{A}\,[A\prec B]} + (1{-}q)\, E_{a,b,f|y,z}^{\tilde{A}\,[B\prec A]} \qquad \forall\, b,f,y,z, \label{eq:cstr_Eabfyz_csep}
\end{align}
with $E_{a,b,f|y,z}^{\tilde{A}\,[A\prec B]}, E_{a,b,f|y,z}^{\tilde{A}\,[B\prec A]}$ satisfying Eqs.~\eqref{eq:cstr_Eabfyz_AprecB}--\eqref{eq:cstr_Eabfyz_BprecA} above, for some $E_{a,b|y}^{\tilde{A}\,[A\prec B]}, E_a^{\tilde{A}_I\,[A\prec B]}, E_{a,b|y}^{\tilde{A}_I\,[B\prec A]} \ge 0$.

\subsubsection{Certifying any (2+$F$)-partite TUU-noncausal process matrix}
\label{app:anyTUU}

From Ref.~\cite{bavaresco19}, a ``TUU-assemblage'' is a set of PSD matrices $(w_{b,f|y,z}^A)_{b,f,y,z}$, with each $w_{b,f|y,z}^A \in \L(\HS^A)$, such that
\begin{align}
\sum_f w_{b,f|y,z}^A = w_{b|y}^A \qquad & \forall\,b,y,z, \label{eq:def_TUU_assemblage1} \\
\sum_b  w_{b|y}^A = \rho_y^{A_I}\otimes\id^{A_O} \qquad & \forall\,y, \label{eq:def_TUU_assemblage2}
\end{align}
for some (PSD) matrices $w_{b|y}^A \in \L(\HS^A)$ and some (normalised) density matrices $\rho_{y}^{A_I} \in \L(\HS^{A_I})$.

\medskip

A TUU-assemblage $(w_{b,f|y,z}^A)_{b,f,y,z}$ is compatible with the order $A\prec B$ if $\rho_{y}^{A_I} = \rho^{A_I}$ does not depend on $y$; it is compatible with the order $B\prec A$ if $w_{b|y}^A$ has the form $w_{b|y}^A = w_{b|y}^{A_I} \otimes \id^{A_O}$ (which then automatically implies the form of Eq.~\eqref{eq:def_TUU_assemblage2}, except for the normalisation). A TUU-assemblage $(w_{b,f|y,z}^A)_{b,f,y,z}$ that can be written as a convex mixture of a TUU-assemblage $(w_{b,f|y,z}^{A\,[A\prec B]})_{b,f,y,z}$ compatible with the order $A\prec B$ and a TUU-assemblage $(w_{b,f|y,z}^{A\,[B\prec A]})_{b,f,y,z}$ compatible with the order $B\prec A$ is said to be causal; otherwise, it is noncausal.

More explicitly, a causal TUU-assemblage $(w_{b,f|y,z}^A)_{b,f,y,z}$ can thus be decomposed as
\begin{align}
w_{b,f|y,z}^A = q \, w_{b,f|y,z}^{A\,[A\prec B]} + (1{-}q)\, w_{b,f|y,z}^{A\,[B\prec A]} \quad & \forall\,b,f,y,z, \label{eq:causal_TUU_assemblage1}
\end{align}
with $q\in [0,1]$ and $w_{b,f|y,z}^{A\,[A\prec B]}, w_{b,f|y,z}^{A\,[B\prec A]}\ge 0$ satisfying
\begin{align}
\sum_f w_{b,f|y,z}^{A\,[A\prec B]} = w_{b|y}^{A\,[A\prec B]} \qquad & \forall\,b,y,z, \label{eq:causal_TUU_assemblage2} \\
\sum_b  w_{b|y}^{A\,[A\prec B]} = \rho^{A_I\,[A\prec B]}\otimes\id^{A_O} \qquad & \forall\,y, \label{eq:causal_TUU_assemblage3} \\
\sum_f w_{b,f|y,z}^{A\,[B\prec A]} = w_{b|y}^{A_I\,[B\prec A]}\otimes\id^{A_O} \qquad & \forall\,b,y,z, \label{eq:causal_TUU_assemblage4}
\end{align}
for some matrices $w_{b|y}^{A\,[A\prec B]} \in \L(\HS^A), w_{b|y}^{A_I\,[B\prec A]} \in \L(\HS^{A_I})$ and some (normalised) density matrices $\rho^{A_I\,[A\prec B]} \in \L(\HS^{A_I})$ (and with $\sum_b \Tr[w_{b|y}^{A_I\,[B\prec A]}] = 1$ for all $y$).

\medskip

One can obtain a ``process TUU-assemblage'' from a (2+$F$)-partite process matrix $W^{ABF}$, some instruments $(M_{b|y}^B)_b$ for Bob (with classical inputs $y$) and some POVMs $(M_{f|z}^F)_f$ for Fiona (with classical inputs $z$), by defining the matrices $w_{b,f|y,z}^A = (M_{b|y}^B\otimes M_{f|z}^F) * W^{ABF}$ (which indeed satisfy Eqs.~\eqref{eq:def_TUU_assemblage1}--\eqref{eq:def_TUU_assemblage2}).%
\footnote{We note that contrary to the TTU case, it remains an open question, whether any TUU-assemblage can be obtained as a process TUU-assemblage~\cite{bavaresco19}.}
A (necessarily causally nonseparable) process matrix that can generate a noncausal TUU-assemblage (as above) is said to be ``TUU-noncausal''. 

\medskip

Consider once again some quantum input spaces $\HS^{\tilde{A}_I}, \HS^{\tilde{A}_O}$ for Alice that are isomorphic to $\HS^{A_I}, \HS^{A_O}$, resp., and the CP map $M_0^{\tilde{A}A}$ of Eq.~\eqref{eq:id_instruments}. Similarly to the previous cases, we find that the induced D-POVM elements $E_{0,b,f|y,z}^{\tilde{A}} = (M_0^{\tilde{A}A}\otimes M_{b|y}^B\otimes M_{f|z}^F)*W^{ABF} = M_0^{\tilde{A}A}*w_{b,f|y,z}^A$ are formally the same, up to a normalisation factor $\frac{1}{d_{A_I}}$, as the matrices $w_{b,f|y,z}^A$ of the process TUU-assemblage obtained as above, but written in the spaces $\tilde{A}_I$ and $\tilde{A}_O$.

Suppose now that the TUU-assemblage $(w_{b,f|y,z}^A)_{b,f,y,z}$ is noncausal---i.e., that $W^{ABF}$ is TUU-noncausal---so that it cannot be decomposed as in Eqs.~\eqref{eq:causal_TUU_assemblage1}--\eqref{eq:causal_TUU_assemblage4} for any PSD matrices $w_{b,f|y,z}^{A\,[A\prec B]}, w_{b,f|y,z}^{A\,[B\prec A]}, w_{b|y}^{A\,[A\prec B]}, \rho^{A_I\,[A\prec B]}, w_{b|y}^{A_I\,[B\prec A]}$.
Translating this onto the $E_{0,b,f|y,z}^{\tilde{A}}$'s, this implies that the latter cannot be decomposed as in Eqs.~\eqref{eq:cstr_Eabfyz_AprecB}--\eqref{eq:cstr_Eabfyz_csep}, for any $E_{0,b,f|y,z}^{\tilde{A}\,[A\prec B]}, E_{0,b,f|y,z}^{\tilde{A}\,[B\prec A]}, E_{0,b|y}^{\tilde{A}\,[A\prec B]}, E_0^{\tilde{A}_I\,[A\prec B]}, E_{0,b|y}^{\tilde{A}_I\,[B\prec A]} \ge 0$.

Hence, any TUU-noncausal (2+$F$)-partite process matrix $W^{ABF}$ can generate some subsets of D-POVM elements $(E_{0,b,f|y,z}^{\tilde{A}})_{b,f,y,z}$ that cannot be decomposed as in Eqs.~\eqref{eq:cstr_Eabfyz_AprecB}--\eqref{eq:cstr_Eabfyz_csep}. Once again, verifying that such a decomposition is impossible can be done with similar techniques to causal witnesses (see Sec.~\ref{app:cones_charact}), which can be measured in practice. Here we only need to trust that Alice's instrument is of the form of Eq.~\eqref{eq:cstr_instr_v0}, but we do not need to trust Bob's instrument or Fiona's POVMs at all, so that this certification of causal nonseparability is MDCI for Alice, but fully DI for Bob and Fiona (``MDCI-DI-DI'').

The quantum switch is again an example of a TUU-noncausal process~\cite{bavaresco19}. Its TUU-noncausality---and hence, its causal nonseparability---can be certified in such a MDCI-DI-DI manner, with a binary classical input for Bob and a fixed POVM for Fiona; see also Sec.~\ref{app:QSwitch}.

\section{Characterisation of the cones of causally separable D-POVMs}
\label{app:cones_charact}

A crucial problem raised in this paper is to characterise the sets of (bipartite or (2+$F$)-partite) causally separable D-POVMs, or of bipartite D-POVM elements $E_{a,b}^{\tilde{A}\tilde{B}}$ of the form of Eq.~\eqref{eq:csep_POVM_00} of the main text (for some fixed $a,b$), or of subsets of (2+$F$)-partite D-POVM elements $(E_{a,b,f}^{\tilde{A}\tilde{B}\tilde{F}})_f$ or $(E_{a,b,f|z}^{\tilde{A}\tilde{B}})_{f,z}$ that can be decomposed as in Eqs.~\eqref{eq:cstr_E00f_csep_1}--\eqref{eq:cstr_E00f_csep_2} or Eqs.~\eqref{eq:cstr_E00fz_csep_1}--\eqref{eq:cstr_E00fz_csep_2}, resp. (for some fixed $a,b$ and all $f,z$'s).

It is convenient, for these characterisations, to drop the global normalisation constraints. The sets that we want to characterise are then closed convex cones. We will follow below the approach that was used in Refs.~\cite{araujo15,branciard16a} to characterise the cones of causally separable process matrices, together with their dual cones that contain the ``witnesses of causal nonseparability''. In particular, the dual cones (which we denote with an asterisk $^*$ or with the ``orthogonal'' symbol $^\perp$ in the case of linear spaces) are typically obtained by using the following duality relations that hold for any of the closed convex cones $\mathcal{K}_1, \mathcal{K}_2$ considered below:
\begin{align}
    (\mathcal{K}_1 + \mathcal{K}_2)^* = \mathcal{K}_1^* \cap \mathcal{K}_2^*, \quad (\mathcal{K}_1 \cap \mathcal{K}_2)^* = \mathcal{K}_1^* + \mathcal{K}_2^*, \label{eq:duality_cones}
\end{align}
where `$+$' here denotes the Minkowski sum.

In the last part of this section we then show explicitly how to write, in terms the cones thus characterised, the primal and dual SDP problems that one can solve, in order to check the causal (non)separability of the D-POVMs or D-POVM elements of interest, and construct explicit causal witnesses.

\subsection{Cone of bipartite case causally separable D-POVMs}
\label{app_cones_N2}

A bipartite D-POVM $\mathbb{E}^{\tilde{A}\prec\tilde{B}}$ with $n_A$ and $n_B$ possible outputs $a$ and $b$, resp., that is compatible with the order $\tilde{A}\prec\tilde{B}$ is a set of $n_An_B$ PSD operators $E_{a,b}^{\tilde{A}\prec\tilde{B}}$ that sum to the identity and satisfy $\sum_b E_{a,b}^{\tilde{A}\prec\tilde{B}} = E_a^{\tilde{A}} \otimes \id^{\tilde{B}}$ for all $a$. Dropping the normalisation (and thus only requiring that the sum of all $E_{a,b}^{\tilde{A}\prec\tilde{B}}$'s---or equivalently, here, of all $E_a^{\tilde{A}}$'s---should be proportional to the identity), the cone of (nonnormalised) causally ordered D-POVMs $\mathbb{E}^{\tilde{A}\prec\tilde{B}} = (E_{a,b}^{\tilde{A}\prec\tilde{B}})_{a,b}$ can then be written as
\begin{align}
\E^{\tilde{A}\prec\tilde{B}} = \P^{n_An_B} \cap \L^{\tilde{A}\prec\tilde{B}},
\end{align}
where $\P$ generically denotes the cone of PSD matrices of appropriate dimensions (taken here to the Cartesian power $n_An_B$) and where $\L^{\tilde{A}\prec\tilde{B}}$ is the linear space
\begin{align}
\L^{\tilde{A}\prec\tilde{B}} & = \{(E_{a,b}^{\tilde{A}\prec\tilde{B}})_{a,b} \,|\, \forall\,a, {\textstyle \sum_b} E_{a,b}^{\tilde{A}\prec\tilde{B}} = E_a^{\tilde{A}} \otimes \id^{\tilde{B}}; \notag \\[-1mm]
& \hspace{44mm} {\textstyle \sum_a} E_a^{\tilde{A}} \propto \id^{\tilde{A}} \}.
\end{align}

The cone of D-POVMs $\mathbb{E}^{\tilde{B}\prec\tilde{A}}$ is obtained in a similar, symmetric manner.
According to Eq.~\eqref{eq:csep_POVM_2} in Definition~\ref{def:csep_POVM_2} of the main text, the cone of (nonnormalised) bipartite causally separable D-POVMs $\mathbb{E}^{\tilde{A}\tilde{B}} = (E_{a,b}^{\tilde{A}\tilde{B}})_{a,b}$ is then obtained as the Minkowski sum
\begin{align}
\E^\text{sep} = \E^{\tilde{A}\prec\tilde{B}} + \E^{\tilde{B}\prec\tilde{A}}.
\end{align}

\medskip

To construct its dual cone, we first need to specify a scalar product. Here we take the one inherited from the (scalar) link product, $\mathbb{S}^{\tilde{A}\tilde{B}} * \mathbb{E}^{\tilde{A}\tilde{B}} = \sum_{a,b} \Tr[ (S_{a,b}^{\tilde{A}\tilde{B}})^T E_{a,b}^{\tilde{A}\tilde{B}}]$ for $\mathbb{S}^{\tilde{A}\tilde{B}} = (S_{a,b}^{\tilde{A}\tilde{B}})_{a,b}$ and $\mathbb{E}^{\tilde{A}\tilde{B}} = (E_{a,b}^{\tilde{A}\tilde{B}})_{a,b}$, so that the dual of a cone $\E$ is the cone $\S = \{\mathbb{S}^{\tilde{A}\tilde{B}}\,|\,\forall \mathbb{E}^{\tilde{A}\tilde{B}}\in\E,\, \mathbb{S}^{\tilde{A}\tilde{B}} * \mathbb{E}^{\tilde{A}\tilde{B}} \ge 0\}$. It is easily verified that $\P^{n_An_B}$ is self-dual, and that%
\footnote{This can be seen by writing $\L^{\tilde{A}\prec\tilde{B}} = \L^{\tilde{A}\prec\tilde{B}}_{\forall a} \cap \L^{\tilde{A}\prec\tilde{B}}_{\propto\id}$ with $\L^{\tilde{A}\prec\tilde{B}}_{\forall a} = \{(E_{a,b}^{\tilde{A}\prec\tilde{B}})_{a,b} | \forall\,a, {\textstyle \sum_b} E_{a,b}^{\tilde{A}\prec\tilde{B}} = E_a^{\tilde{A}} \otimes \id^{\tilde{B}} \}$ and $\L^{\tilde{A}\prec\tilde{B}}_{\propto\id} = \{(E_{a,b}^{\tilde{A}\prec\tilde{B}})_{a,b} | \sum_{a,b} E_{a,b}^{\tilde{A}\prec\tilde{B}} \propto \id^{\tilde{A}\tilde{B}} \}$, such that $(\L^{\tilde{A}\prec\tilde{B}}_{\forall a})^\perp = \{(S_{a,b}^{\tilde{A}\tilde{B}} = S_a^{\tilde{A}\prec\tilde{B}})_{a,b} | \forall\,a, \Tr_{\tilde{B}}[S_a^{\tilde{A}\prec\tilde{B}}] = 0 \}$ and $(\L^{\tilde{A}\prec\tilde{B}}_{\propto\id})^\perp = \{(S_{a,b}^{\tilde{A}\tilde{B}} = S^{\tilde{A}\prec\tilde{B}})_{a,b} | \Tr[S^{\tilde{A}\prec\tilde{B}}] = 0 \}$, and then using Eq.~\eqref{eq:duality_cones}.}
$(\L^{\tilde{A}\prec\tilde{B}})^* = (\L^{\tilde{A}\prec\tilde{B}})^\perp = \{(S_{a,b}^{\tilde{A}\tilde{B}} = S_a^{\tilde{A}\prec\tilde{B}} + S^{\tilde{A}\prec\tilde{B}})_{a,b} \,|\, \forall\,a, \Tr_{\tilde{B}}[S_a^{\tilde{A}\prec\tilde{B}}] = 0; \Tr[S^{\tilde{A}\prec\tilde{B}}] = 0 \}$.
Using the duality relations of Eq.~\eqref{eq:duality_cones}, we thus obtain
\begin{align}
(\E^\text{sep})^* = (\E^{\tilde{A}\prec\tilde{B}})^* \cap (\E^{\tilde{B}\prec\tilde{A}})^*
\label{eq:dual_cone_2}
\end{align}
with
\begin{align}
(\E^{\tilde{A}\prec\tilde{B}})^* & = \P^{n_An_B} + (\L^{\tilde{A}\prec\tilde{B}})^\perp , \notag \\
& = \{(S_{a,b}^{\tilde{A}\tilde{B}} = S_{\text{PSD};a,b}^{\tilde{A}\prec\tilde{B}} + S_a^{\tilde{A}\prec\tilde{B}} + S^{\tilde{A}\prec\tilde{B}})_{a,b} \notag \\[-1mm]
& \hspace{7mm} |\, \forall\,a,b,\, S_{\text{PSD};a,b}^{\tilde{A}\prec\tilde{B}} \ge 0, \Tr_{\tilde{B}}[S_a^{\tilde{A}\prec\tilde{B}}] = 0, \notag \\[-1mm]
& \hspace{39mm} \Tr[S^{\tilde{A}\prec\tilde{B}}] = 0 \}
\label{eq:dual_cone_2_2}
\end{align}
and similarly for $(\E^{\tilde{B}\prec\tilde{A}})^*$.

Any set of operators $\mathbb{S}^{\tilde{A}\tilde{B}} = (S_{a,b}^{\tilde{A}\tilde{B}})_{a,b}$ in the dual cone $(\E^\text{sep})^*$ acts as a ``witness of causal nonseparability'' for D-POVMs, in the sense that by definition all causally separable D-POVMs satisfy $\mathbb{S}^{\tilde{A}\tilde{B}} * \mathbb{E}^{\tilde{A}\tilde{B}} \ge 0$---hence, if one gets a value $\mathbb{S}^{\tilde{A}\tilde{B}} * \mathbb{E}^{\tilde{A}\tilde{B}} < 0$, this certifies that the D-POVM is causally nonseparable. Note, furthermore, that since the set of causally separable D-POVMs is closed and convex, then by the separating hyperplane theorem~\cite{rockafellar70}, for any causally nonseparable D-POVM there exists a witness that certifies it. A witness can be measured in practice, as clarified around Eq.~\eqref{eq:reconstruct_witness} in the main text.

\subsection{Cone of (2+$F$)-partite causally separable D-POVMs}

The characterisation of the cone of (2+$F$)-partite causally separable D-POVMs is similar to that in the bipartite case. We start by characterising the cone D-POVMs $\mathbb{E}^{\tilde{A}\prec\tilde{B}\prec\tilde{F}}$ (with $n_A$, $n_B$ and $n_F$ possible outputs $a$, $b$ and $f$, resp.) compatible with the order $\tilde{A}\prec\tilde{B}\prec\tilde{F}$, as
\begin{align}
\E^{\tilde{A}\prec\tilde{B}\prec\tilde{F}} = \P^{n_An_Bn_F} \cap \L^{\tilde{A}\prec\tilde{B}\prec\tilde{F}},
\end{align}
with
\begin{align}
& \L^{\tilde{A}\prec\tilde{B}\prec\tilde{F}} \notag \\
& = \{(E_{a,b,f}^{\tilde{A}\prec\tilde{B}\prec\tilde{F}})_{a,b,f} \,|\, \forall\,a,b, {\textstyle \sum_f} E_{a,b,f}^{\tilde{A}\prec\tilde{B}\prec\tilde{F}} = E_{a,b}^{\tilde{A}\prec\tilde{B}} \otimes \id^{\tilde{F}}; \notag \\
& \hspace{12mm} \forall\,a, {\textstyle \sum_b} E_{a,b}^{\tilde{A}\prec\tilde{B}} = E_a^{\tilde{A}} \otimes \id^{\tilde{B}}; {\textstyle \sum_a} E_a^{\tilde{A}} \propto \id^{\tilde{A}} \}.
\end{align}

The cone of D-POVMs $\mathbb{E}^{\tilde{B}\prec\tilde{A}\prec\tilde{F}}$ is obtained in a similar way, and according to Eq.~\eqref{eq:csep_POVM_2F} in Definition~\ref{def:csep_POVM_2F}, the cone of (2+$F$)-partite causally separable D-POVMs $\mathbb{E}^{\tilde{A}\tilde{B}\tilde{F}} = (E_{a,b,f}^{\tilde{A}\tilde{B}\tilde{F}})_{a,b,f}$ is
\begin{align}
\E^\text{sep} = \E^{\tilde{A}\prec\tilde{B}\prec\tilde{F}} + \E^{\tilde{B}\prec\tilde{A}\prec\tilde{F}}.
\end{align}

Its dual cone (of ``witnesses of causal nonseparability'' $\mathbb{S}^{\tilde{A}\tilde{B}\tilde{F}} = (S_{a,b,f}^{\tilde{A}\tilde{B}\tilde{F}})_{a,b,f}$, now for the scalar product $\mathbb{S}^{\tilde{A}\tilde{B}\tilde{F}} * \mathbb{E}^{\tilde{A}\tilde{B}\tilde{F}} = \sum_{a,b,f} \Tr[ (S_{a,b,f}^{\tilde{A}\tilde{B}\tilde{F}})^T E_{a,b,f}^{\tilde{A}\tilde{B}\tilde{F}}]$) is then
\begin{align}
(\E^\text{sep})^* = (\E^{\tilde{A}\prec\tilde{B}\prec\tilde{F}})^* \cap (\E^{\tilde{B}\prec\tilde{A}\prec\tilde{F}})^* \label{eq:dual_cone_2F}
\end{align}
with 
\begin{align}
& \!\!(\E^{\tilde{A}\prec\tilde{B}\prec\tilde{F}})^* = \P^{n_An_Bn_F} + (\L^{\tilde{A}\prec\tilde{B}\prec\tilde{F}})^\perp \notag \\
& \!\!\!\!\!\! = \! \{(S_{a,b,f}^{\tilde{A}\tilde{B}\tilde{F}} \!=\! S_{\text{PSD};a,b,f}^{\tilde{A}\prec\tilde{B}\prec\tilde{F}} \!+\! S_{a,b}^{\tilde{A}\prec\tilde{B}\prec\tilde{F}} \!\!+\! S_a^{\tilde{A}\prec\tilde{B}\prec\tilde{F}} \!\!+\! S^{\tilde{A}\prec\tilde{B}\prec\tilde{F}})_{a,b,f} \notag \\
& \hspace{7mm} |\, \forall\,a,b,f,\, S_{\text{PSD};a,b,f}^{\tilde{A}\prec\tilde{B}\prec\tilde{F}}  \ge 0, \Tr_{\tilde{F}}[S_{a,b}^{\tilde{A}\prec\tilde{B}\prec\tilde{F}}] = 0, \notag \\
& \hspace{14mm} \Tr_{\tilde{B}\tilde{F}}[S_a^{\tilde{A}\prec\tilde{B}\prec\tilde{F}}] = 0, \Tr[S^{\tilde{A}\prec\tilde{B}\prec\tilde{F}}] = 0 \} \label{eq:dual_cone_2F_2}
\end{align}
and similarly for $(\E^{\tilde{B}\prec\tilde{A}\prec\tilde{F}})^*$.

\subsection{Cone of bipartite single D-POVM elements \texorpdfstring{$E_{0,0}^{\tilde{A}\tilde{B}}$}{$E_{0,0}^{AB}$} of the form of Eq.~\eqref{eq:csep_POVM_00}}

As we saw, in the case where we impose the structure of Eq.~\eqref{eq:cstr_instr_v0} for Alice and Bob's instruments, the causal separability of a bipartite process matrix $W^{AB}$ implies that each individual D-POVM element $E_{a,b}^{\tilde{A}\tilde{B}}$---e.g., $E_{0,0}^{\tilde{A}\tilde{B}}$ for $a=b=0$---is of the form of Eq.~\eqref{eq:csep_POVM_00} in the main text. Once again dropping the global normalisation, the cone of matrices of this form is simply obtained as%
\footnote{The subscripts `$0,0$' in the names of the cones (e.g., in $\E_{0,0}^\text{sep}$) indicate that these are cones of D-POVM elements corresponding to some fixed $a,b$ (say, $a=b=0$), as opposed to the full sets of D-POVM elements for all $a,b$. In $\E_{0,0,\cdot}^\text{sep}$ further below the dot indicates that we still consider all possible values $f$.}
\begin{align}
\E_{0,0}^\text{sep} = \E_{0,0}^{\tilde{A}\prec\tilde{B}} + \E_{0,0}^{\tilde{B}\prec\tilde{A}}
\end{align}
with 
\begin{align}
\E_{0,0}^{\tilde{A}\prec\tilde{B}} & = \{ E_{0,0}^{\tilde{A}\prec\tilde{B}_I} \otimes \id^{\tilde{B}_O} \,|\, E_{0,0}^{\tilde{A}\prec\tilde{B}_I} \ge 0 \}, \label{eq:cone_E00_ABI}
\end{align}
and similarly for $\E_{0,0}^{\tilde{B}\prec\tilde{A}}$.

\medskip

Its dual cone (of ``witnesses'' $S_{0,0}^{\tilde{A}\tilde{B}}$, for the scalar link product $S_{0,0}^{\tilde{A}\tilde{B}} * E_{0,0}^{\tilde{A}\tilde{B}} = \Tr[ (S_{0,0}^{\tilde{A}\tilde{B}})^T E_{0,0}^{\tilde{A}\tilde{B}}]$), is
\begin{align}
(\E_{0,0}^\text{sep})^* = (\E_{0,0}^{\tilde{A}\prec\tilde{B}})^* \cap (\E_{0,0}^{\tilde{B}\prec\tilde{A}})^*
\end{align}
with 
\begin{align}
(\E_{0,0}^{\tilde{A}\prec\tilde{B}})^* & = \{ S_{0,0}^{\tilde{A}\tilde{B}} \,|\, \Tr_{\tilde{B}_O} [S_{0,0}^{\tilde{A}\tilde{B}}] \ge 0 \},
\end{align}
and similarly for $(\E_{0,0}^{\tilde{B}\prec\tilde{A}})^*$.

\medskip

Let us note here that the cone $\E_{0,0}^\text{sep}$ above is not simply isomorphic to the cone $\W^\text{sep}$ of (nonnormalised) causally separable process matrices, characterised e.g.\ in Refs.~\cite{araujo15,branciard16a}. Indeed, as noted in Sec.~\ref{app:subsubsec:Cstr_each_Eab}, the operators $E_{0,0}^{\tilde{A}\prec\tilde{B}_I}$ in Eq.~\eqref{eq:cone_E00_ABI} are not bound to satisfy $_{[1-\tilde{A}_O]\tilde{B}_I}E_{0,0}^{\tilde{A}\prec\tilde{B}_I}=0$, contrary to the process matrices $W^{A\prec B_I}$ in the decomposition of a causally separable process, that must satisfy $_{[1-A_O]B_I}W^{A\prec B_I}=0$. The cone $\E_{0,0}^\text{sep}$ is in this sense ``larger'' than $\W^\text{sep}$, and its dual cone $(\E_{0,0}^\text{sep})^*$ is correspondingly ``smaller'' than the cone $(\W^\text{sep})^*$ of standard causal witnesses~\cite{araujo15,branciard16a}. Thus, not all standard causal witnesses can be translated into a causal witness for D-POVM elements $E_{0,0}^{\tilde{A}\tilde{B}}$;%
\footnote{E.g., a standard causal witness $S^{AB}$ (satisfying $S^{AB} * W^{AB} \ge 0$ for any causally separable $W^{AB}$) could still satisfy $S^{AB} * (\underline{W}^{AB_I}\otimes \id^{B_O}) < 0$ for some PSD operator $\underline{W}^{AB_I}$ such that $_{[1-A_O]B_I}\underline{W}^{AB_I}\neq 0$ (hence not a valid process matrix), and would thus not translate into a witness for D-POVM elements $E_{0,0}^{\tilde{A}\tilde{B}}$.}
instead, our claim is that for any causally nonseparable $W^{AB}$, \emph{there exists} a causal witness that detects it, and that can be turned into a causal witness for the corresponding $E_{0,0}^{\tilde{A}\tilde{B}}$.

\subsection{Cone of subsets of (2+$F$)-partite D-POVM elements \texorpdfstring{$(E_{0,0,f}^{\tilde{A}\tilde{B}\tilde{F}})_f$}{$(E_{0,0,f}^{ABF})_f$}  satisfying Eqs.~\eqref{eq:cstr_E00f_csep_1}--\eqref{eq:cstr_E00f_csep_2}}

Imposing again the structure of Eq.~\eqref{eq:cstr_instr_v0} for Alice and Bob's instruments in the (2+$F$)-partite case, the causal separability of $W^{ABF}$ implies that the subset of D-POVM elements $(E_{a,b,f}^{\tilde{A}\tilde{B}\tilde{F}})_f$, for some fixed $a,b$ and all $f$'s---e.g., $(E_{0,0,f}^{\tilde{A}\tilde{B}\tilde{F}})_f$ for $a=b=0$---satisfies Eqs.~\eqref{eq:cstr_E00f_csep_1}--\eqref{eq:cstr_E00f_csep_2}. The cone of such subsets of D-POVM elements (for $n_F$ possible values of $f$) is obtained as
\begin{align}
\E_{0,0,\cdot}^\text{sep} = \E_{0,0,\cdot}^{\tilde{A}\prec\tilde{B}\prec\tilde{F}} + \E_{0,0,\cdot}^{\tilde{B}\prec\tilde{A}\prec\tilde{F}}
\end{align}
with 
\begin{align}
\E_{0,0,\cdot}^{\tilde{A}\prec\tilde{B}\prec\tilde{F}} & = \P^{n_F} \cap \L_{0,0,\cdot}^{\tilde{A}\prec\tilde{B}\prec\tilde{F}}, \notag \\
\L_{0,0,\cdot}^{\tilde{A}\prec\tilde{B}\prec\tilde{F}} \! & = \! \{ (E_{0,0,f}^{\tilde{A}\prec\tilde{B}\prec\tilde{F}})_f \,|\, {\textstyle \sum_f} E_{0,0,f}^{\tilde{A}\prec\tilde{B}\prec\tilde{F}}{=}E_{0,0}^{\tilde{A}\prec\tilde{B}_I} \!\otimes\! \id^{\tilde{B}_O\tilde{F}} \},
\end{align}
and similarly for $\E_{0,0,\cdot}^{\tilde{B}\prec\tilde{A}\prec\tilde{F}}$.

\medskip

Its dual cone (of ``witnesses'' $(S_{a,b,f}^{\tilde{A}\tilde{B}\tilde{F}})_f$, for the scalar product $(S_{a,b,f}^{\tilde{A}\tilde{B}\tilde{F}})_f * (E_{a,b,f}^{\tilde{A}\tilde{B}\tilde{F}})_f = \sum_f \Tr[ (S_{a,b,f}^{\tilde{A}\tilde{B}\tilde{F}})^T E_{a,b,f}^{\tilde{A}\tilde{B}\tilde{F}}]$), is
\begin{align}
(\E_{0,0,\cdot}^\text{sep})^* = (\E_{0,0,\cdot}^{\tilde{A}\prec\tilde{B}\prec\tilde{F}})^* \cap (\E_{0,0,\cdot}^{\tilde{B}\prec\tilde{A}\prec\tilde{F}})^*
\end{align}
with 
\begin{align}
& (\E_{0,0,\cdot}^{\tilde{A}\prec\tilde{B}\prec\tilde{F}})^* = \P^{n_F} + (\L_{0,0,\cdot}^{\tilde{A}\prec\tilde{B}\prec\tilde{F}})^\perp \notag \\
& = \{ (S_{0,0,f}^{\tilde{A}\tilde{B}\tilde{F}} = S_{\text{PSD};f}^{\tilde{A}\prec\tilde{B}\prec\tilde{F}} + S^{\tilde{A}\prec\tilde{B}\prec\tilde{F}})_f \notag \\
& \hspace{10mm} |\, \forall f, S_{\text{PSD};f}^{\tilde{A}\prec\tilde{B}\prec\tilde{F}} \ge 0, \Tr_{\tilde{B}_OF} [S^{\tilde{A}\prec\tilde{B}\prec\tilde{F}}] = 0 \},
\end{align}
and similarly for $(\E_{0,0,\cdot}^{\tilde{B}\prec\tilde{A}\prec\tilde{F}})^*$.

\subsection{(2+$F$)-partite case with classical inputs for Fiona: cone of subsets of D-POVM elements \texorpdfstring{$(E_{0,0,f|z}^{\tilde{A}\tilde{B}})_{f,z}$}{$(E_{0,0,f|z}^{AB})_{f,z}$} satisfying Eqs.~\eqref{eq:cstr_E00fz_csep_1}--\eqref{eq:cstr_E00fz_csep_2}}

In the (2+$F$)-partite case where Fiona now has classical inputs (and Alice and Bob's instruments are still of the form of Eq.~\eqref{eq:cstr_instr_v0}), the subsets of D-POVM elements $(E_{a,b,f|z}^{\tilde{A}\tilde{B}})_{f,z}$, for some fixed $a,b$ and all $f,z$'s---e.g., $(E_{0,0,f|z}^{\tilde{A}\tilde{B}})_{f,z}$ for $a=b=0$---induced by a causally separable process matrix $W^{ABF}$ satisfy Eqs.~\eqref{eq:cstr_E00fz_csep_1}--\eqref{eq:cstr_E00fz_csep_2}. The cone of such subsets of D-POVM elements (for $n_\sharp$ possible pairs $(f,z)$) is now
\begin{align}
\E_{0,0,\cdot|\cdot}^\text{sep} = \E_{0,0,\cdot|\cdot}^{\tilde{A}\prec\tilde{B}} + \E_{0,0,\cdot|\cdot}^{\tilde{B}\prec\tilde{A}}
\end{align}
with 
\begin{align}
\E_{0,0,\cdot|\cdot}^{\tilde{A}\prec\tilde{B}} & = \P^{n_\sharp} \cap \L_{0,0,\cdot|\cdot}^{\tilde{A}\prec\tilde{B}} , \notag \\
\L_{0,0,\cdot|\cdot}^{\tilde{A}\prec\tilde{B}} & = \{(E_{0,0,f|z}^{\tilde{A}\prec\tilde{B}})_{f,z} | \forall\,z, {\textstyle \sum_f} E_{0,0,f|z}^{\tilde{A}\prec\tilde{B}} = E_{0,0}^{\tilde{A}\prec\tilde{B}_I} \otimes \id^{\tilde{B}_O} \},
\end{align}
and similarly for $\E_{0,0,\cdot|\cdot}^{\tilde{B}\prec\tilde{A}}$.

\medskip

Its dual cone (of ``witnesses'' $(S_{0,0,f|z}^{\tilde{A}\tilde{B}})_{f,z}$, for the scalar product $(S_{0,0,f|z}^{\tilde{A}\tilde{B}})_{f,z} * (E_{0,0,f|z}^{\tilde{A}\tilde{B}})_{f,z} = \sum_{f,z} \Tr[ (S_{0,0,f|z}^{\tilde{A}\tilde{B}})^T E_{0,0,f|z}^{\tilde{A}\tilde{B}}]$), is
\begin{align}
(\E_{0,0,\cdot|\cdot}^\text{sep})^* = (\E_{0,0,\cdot|\cdot}^{\tilde{A}\prec\tilde{B}})^* \cap (\E_{0,0,\cdot|\cdot}^{\tilde{B}\prec\tilde{A}})^*
\end{align}
with 
\begin{align}
& (\E_{0,0,\cdot|\cdot}^{\tilde{A}\prec\tilde{B}})^* = \P^{n_\sharp}  + (\L_{0,0,\cdot|\cdot}^{\tilde{A}\prec\tilde{B}})^\perp , \notag \\
& = \{(S_{0,0,f|z}^{\tilde{A}\tilde{B}} = S_{\text{PSD};f,z}^{\tilde{A}\prec\tilde{B}} + S_z^{\tilde{A}\prec\tilde{B}})_{f,z} \notag \\
& \hspace{8mm} |\, \forall\, f,z,\, S_{\text{PSD};f,z}^{\tilde{A}\prec\tilde{B}} \ge 0, {\textstyle \sum_z} \Tr_{\tilde{B}_O}[S_z^{\tilde{A}\prec\tilde{B}}] = 0 \},
\end{align}
and similarly for $(\E_{0,0,\cdot|\cdot}^{\tilde{B}\prec\tilde{A}})^*$.

\subsection{(2+$F$)-partite case with classical inputs for Bob and Fiona: cone of subsets of D-POVM elements \texorpdfstring{$(E_{0,b,f|y,z}^{\tilde{A}})_{b,f,y,z}$}{$(E_{0,b,f|y,z}^A)_{b,f,y,z}$} satisfying Eqs.~\eqref{eq:cstr_Eabfyz_AprecB}--\eqref{eq:cstr_Eabfyz_csep}}

In the (2+$F$)-partite case where Bob and Fiona both have classical inputs (and Alice's instrument is still of the form of Eq.~\eqref{eq:cstr_instr_v0}), the subsets of D-POVM elements $(E_{a,b,f|y,z}^{\tilde{A}})_{b,f,y,z}$, for some fixed $a$ and all $b,f,y,z$'s---e.g., $(E_{0,b,f|y,z}^{\tilde{A}})_{b,f,y,z}$ for $a=0$---induced by a causally separable process matrix $W^{ABF}$ satisfy Eqs.~\eqref{eq:cstr_Eabfyz_AprecB}--\eqref{eq:cstr_Eabfyz_csep}. The cone of such subsets of D-POVM elements (for $n_\sharp$ possible values of $(b,f,y,z)$) is now
\begin{align}
\E_{0,\cdot,\cdot|\cdot,\cdot}^\text{sep} = \E_{0,\cdot,\cdot|\cdot,\cdot}^{\tilde{A}\,[A\prec B]} + \E_{0,\cdot,\cdot|\cdot,\cdot}^{\tilde{A}\,[B\prec A]}
\end{align}
with 
\begin{align}
\E_{0,\cdot,\cdot|\cdot,\cdot}^{\tilde{A}\,[A\prec B]} & = \P^{n_\sharp} \!\cap\! \L_{0,\cdot,\cdot|\cdot,\cdot}^{\tilde{A}\,[A\prec B]} , \ \ \E_{0,\cdot,\cdot|\cdot,\cdot}^{\tilde{A}\,[B\prec A]} = \P^{n_\sharp} \!\cap\! \L_{0,\cdot,\cdot|\cdot,\cdot}^{\tilde{A}\,[B\prec A]} , \notag \\[1mm]
\L_{0,\cdot,\cdot|\cdot,\cdot}^{\tilde{A}\,[A\prec B]} & = \{(E_{0,b,f|y,z}^{\tilde{A}\,[A\prec B]})_{b,f,y,z} \notag \\[-1mm]
& \qquad |\, \forall\,b,y,z, {\textstyle \sum_f} E_{0,b,f|y,z}^{\tilde{A}\,[A\prec B]} = E_{0,b|y}^{\tilde{A}\,[A\prec B]}, \notag \\[-1mm]
& \qquad \ \, \forall\,y, {\textstyle \sum_b} E_{0,b|y}^{\tilde{A}\,[A\prec B]} = E_0^{\tilde{A}_I\,[A\prec B]}\otimes\id^{\tilde{A}_O} \}, \notag \\[1mm]
\L_{0,\cdot,\cdot|\cdot,\cdot}^{\tilde{A}\,[B\prec A]} & = \{(E_{0,b,f|y,z}^{\tilde{A}\,[B\prec A]})_{b,f,y,z} \notag \\[-1mm]
& \quad\ \ |\, \forall\,b,y,z, {\textstyle \sum_f} E_{0,b,f|y,z}^{\tilde{A}\,[B\prec A]} = E_{0,b|y}^{\tilde{A}_I\,[B\prec A]}\otimes\id^{\tilde{A}_O} \}.
\end{align}

\medskip

Its dual cone (of ``witnesses'' $(S_{0,b,f|y,z}^{\tilde{A}})_{b,f,y,z}$, for the scalar product $(S_{0,b,f|y,z}^{\tilde{A}})_{b,f,y,z} * (E_{0,b,f|y,z}^{\tilde{A}})_{b,f,y,z} = \sum_{b,f,y,z} \Tr[ (S_{0,b,f|y,z}^{\tilde{A}})^T E_{0,b,f|y,z}^{\tilde{A}}]$), is
\begin{align}
(\E_{0,\cdot,\cdot|\cdot,\cdot}^\text{sep})^* = (\E_{0,\cdot,\cdot|\cdot,\cdot}^{\tilde{A}\,[A\prec B]})^* \cap (\E_{0,\cdot,\cdot|\cdot,\cdot}^{\tilde{A}\,[B\prec A]})^*
\end{align}
with
\begin{align}
& (\E_{0,\cdot,\cdot|\cdot,\cdot}^{\tilde{A}\,[A\prec B]})^* = \P^{n_\sharp}  + (\L_{0,\cdot,\cdot|\cdot,\cdot}^{\tilde{A}\,[A\prec B]})^\perp , \notag \\
& = \{(S_{0,b,f|y,z}^{\tilde{A}} = S_{\text{PSD};b,f,y,z}^{\tilde{A}\,[A\prec B]} + S_{b,y,z}^{\tilde{A}\,[A\prec B]})_{b,f,y,z} \notag \\
& \hspace{18mm} |\, \forall\, b,f,y,z,\, S_{\text{PSD};b,f,y,z}^{\tilde{A}\,[A\prec B]} \ge 0, \notag \\
& \hspace{20mm} \forall\,b,y, {\textstyle \sum_z} S_{b,y,z}^{\tilde{A}\,[A\prec B]} = S_y^{\tilde{A}\,[A\prec B]}, \notag \\
& \hspace{20mm} {\textstyle \sum_y} \Tr_{\tilde{A}_O}[S_y^{\tilde{A}\,[A\prec B]}] = 0 \}
\end{align}
and
\begin{align}
& (\E_{0,\cdot,\cdot|\cdot,\cdot}^{\tilde{A}\,[B\prec A]})^* = \P^{n_\sharp}  + (\L_{0,\cdot,\cdot|\cdot,\cdot}^{\tilde{A}\,[B\prec A]})^\perp , \notag \\
& = \{(S_{0,b,f|y,z}^{\tilde{A}} = S_{\text{PSD};b,f,y,z}^{\tilde{A}\,[B\prec A]} + S_{b,y,z}^{\tilde{A}\,[B\prec A]})_{b,f,y,z} \notag \\
& \hspace{18mm} |\, \forall\, b,f,y,z,\, S_{\text{PSD};b,f,y,z}^{\tilde{A}\,[B\prec A]} \ge 0, \notag \\
& \hspace{20mm} \forall\,b,y, {\textstyle \sum_z} \Tr_{\tilde{A}_O}[S_{b,y,z}^{\tilde{A}\,[B\prec A]}] = 0 \}.
\end{align}

\subsection{Constructing witnesses of causal nonseparability via semidefinite programming}
\label{app:SDPs}

Now that the cones of interest have been characterised, one can use these characterisations to check whether a given D-POVM is causally separable or not, or whether some D-POVM elements have a causally separable structure or not. Interestingly, this can be done via semidefinite programming (SDP); this approach also allows one, in the case of causal nonseparability, to construct explicit ``causal witnesses''.

\medskip

Consider an object $\mathbb{E}$ (e.g., here a D-POVM or a D-POVM element), of which we want to test the causal separability, i.e., its membership in some cone $\E^\text{sep}$. A simple way to express this problem is to look at the robustness when mixing $\mathbb{E}$ with some other object $\mathbb{E}^\circ \in \E^\text{sep}$, and solve the SDP \emph{primal problem}
\begin{align}
\min \, r \quad \text{s.t.} \quad \mathbb{E} + r\, \mathbb{E}^\circ \in \E^\text{sep},\ r\ge 0. \label{eq:primal}
\end{align}
If the optimal solution to this problem is found to be $r^* > 0$, then this implies that $\mathbb{E}\notin\E^\text{sep}$ is causally nonseparable; this optimal solution $r^*$ is called the ``robustness'' of $\mathbb{E}$ with respect to $\mathbb{E}^\circ$.

The primal problem~\eqref{eq:primal} is intimately related to its SDP \emph{dual problem}, which takes the form%
\footnote{To get the dual from the primal problem, one may follow the approach presented in Appendix~E of Ref.~\cite{araujo15}, with (in the notations of that paper) $E = \L_{\mathbb{E}} \times \mathbb{R}$ (where $\L_{\mathbb{E}}$ is the linear space in which all possible $\mathbb{E}$'s live), $\K = \E^\text{sep} \times \mathbb{R}^+$ (such that $\K^* = (\E^\text{sep})^* \times \mathbb{R}^+$), $\L = \{(r \mathbb{E}^\circ, r) \,|\, r \in \mathbb{R}\}$ (such that $\L^\perp = \{(\mathbb{S}, -\mathbb{S} * \mathbb{E}^\circ) \,|\, \mathbb{S} \in \L_{\mathbb{E}}'\}$), $b = (\mathbb{E},0)$ and $c = (0,1)$.}
\begin{align}
\min \, \mathbb{S} * \mathbb{E} \quad \text{s.t.} \quad & \mathbb{S} \in (\E^\text{sep})^*, \  \mathbb{S} * \mathbb{E}^\circ \leq 1. \label{eq:dual}
\end{align}
If the optimal solution $\mathbb{S}^*$ of that problem gives $\mathbb{S}^* * \mathbb{E} < 0$, then this provides an explicit witness $\mathbb{S}^*$ certifying the causal nonseparability of $\mathbb{E}$, that can be used and measured in practice.

Let us also recall that provided $\mathbb{E}^\circ$ is taken to be in the interior of $\E^\text{sep}$, so that the assumptions of the ``Duality Theorem'' (cf., e.g., Theorem~8 in Appendix~E of Ref.~\cite{araujo15}) hold, the optimal solutions $r^*$ and $\mathbb{S}^*$ of the primal and dual problems above related through
\begin{align}
r^* = - \mathbb{S}^* * \mathbb{E}.
\end{align}

\medskip

For the purpose of the present paper, $\mathbb{E}$ can be taken to be a D-POVM $\mathbb{E}^{\tilde{A}\tilde{B}} = (E_{a,b}^{\tilde{A}\tilde{B}})_{a,b}$ or $\mathbb{E}^{\tilde{A}\tilde{B}\tilde{F}} = (E_{a,b,f}^{\tilde{A}\tilde{B}\tilde{F}})_{a,b,f}$, or just a D-POVM element $E_{0,0}^{\tilde{A}\tilde{B}}$, or some subsets of D-POVM elements $(E_{0,0,f}^{\tilde{A}\tilde{B}\tilde{F}})_f$, $(E_{0,0,f|z}^{\tilde{A}\tilde{B}})_{f,z}$ or $(E_{0,b,f|y,z}^{\tilde{A}})_{b,f,y,z}$. $\E^\text{sep}$ and $\mathbb{E}^\circ$ are to be taken accordingly (with $\mathbb{E}^\circ$ typically taken to describe some ``fully random'' object, to then provide the ``random robustness'' of $\mathbb{E}$). We then refer to the previous subsections for the characterisations of $\E^\text{sep}$ and $(\E^\text{sep})^*$ that are needed to implement the primal and dual problems above.

\section{Examples}
\label{app:examples}

\subsection{SDI-QI causal nonseparability of Feix \emph{et al.}'s process matrix~\cite{feix16}}
\label{app:Feix}

In Ref.~\cite{feix16}, Feix \emph{et al.} introduced the following family of bipartite process matrices, with two-dimensional (i.e., qubit) input and output spaces for Alice and Bob:
\begin{align}
W_\text{FAB}(q,\epsilon) = \id^\circ & + \frac{q}{12}(\id X X \id + \id Y Y \id + \id Z Z \id ) \notag \\
& + \frac{1-q+\epsilon}{4}Z \id X Z
\end{align}
with $\id^\circ = \id^{AB}/4$, $q\in[0,1]$ and $|1-q+\epsilon| \leq \sqrt{\frac{(1-q)(q+3)}{3}}$ so as to ensure that $W_\text{FAB}(q,\epsilon)$ is PSD (as required for a valid process matrix).
Here $X$, $Y$ and $Z$ are the Pauli matrices; we have omitted tensor products, and have written these in the order $\HS^{A_I}\otimes\HS^{A_O}\otimes\HS^{B_I}\otimes\HS^{B_O}$ (e.g., $\id X X \id = \id^{A_I}\otimes X^{A_O}\otimes X^{B_I}\otimes \id^{B_O}$).
It was shown in Ref.~\cite{feix16} that $W_\text{FAB}(q,\epsilon)$ is causally nonseparable for $\epsilon > 0$, in which case its random robustness~\cite{araujo15,branciard16a} is $\epsilon$. In the following we will take the values $q = \sqrt{3}-1$ and $\epsilon = \frac{4}{\sqrt{3}}-2$, which give the maximal random robustness $\epsilon = \frac{4}{\sqrt{3}}-2\simeq 0.309$, and simply write $W_\text{FAB} = W_\text{FAB}(q,\epsilon)$ for those values.

Mixing this process matrix with fully white noise, described by the process matrix $\id^\circ$, we then define
\begin{align}
W_\text{FAB}(r) = \frac{1}{1+r}(W_\text{FAB} + r \, \id^\circ)
\end{align}
for $r\ge 0$.
The smallest value of $r$ such that $W_\text{FAB}(r)$ is causally separable defines the random robustness of $W_\text{FAB}$~\cite{araujo15,branciard16a}: as just recalled, $W_\text{FAB}(r)$ is thus causally nonseparable for all $r \le \frac{4}{\sqrt{3}}-2\simeq 0.309$.

\medskip

In our scenario with trusted quantum inputs, we found that this process can generate a causally nonseparable D-POVM for all $r\lesssim 0.113$.

To obtain this, we considered a 1-qubit input state for Alice (isomorphic to $\HS^{A_O}$) and a 2-qubit input state for Bob (with a state space decomposing as $\HS^{\tilde{B}} = \HS^{\tilde{B}_I\tilde{B}_O}$, with $\HS^{\tilde{B}_I}, \HS^{\tilde{B}_O}$ isomorphic to $\HS^{B_I}, \HS^{B_O}$, resp.), and the instruments $(M_a^{\tilde{A}A})_a$, $(M_b^{\tilde{B}B})_b$ defined by
\begin{align}
M_a^{\tilde{A}A} & = \ketbra{a}{a}^{A_I} \otimes \dketbra{\id}{\id}^{\tilde{A}A_O}, \notag \\
M_0^{\tilde{B}B} & = \ketbra{\psi}{\psi}^{\tilde{B}_IB_I} \otimes (\ketbra{0,0}{0,0} + \ketbra{1,1}{1,1})^{\tilde{B}_OB_O}, \notag \\
M_1^{\tilde{B}B} & = \id^{\tilde{B}_IB_I} \otimes (\ketbra{0,0}{0,0} + \ketbra{1,1}{1,1})^{\tilde{B}_OB_O} \notag \\
& \quad - \ketbra{+}{+}^{\tilde{B}_I} \otimes \ketbra{-}{-}^{B_I} \otimes Z^{\tilde{B}_O} \otimes Z^{B_O} - M_0^{\tilde{B}B} \label{eq:instr_W_FAB}
\end{align}
with $\ket{\psi} = (\ket{+,+} + \xi \ket{-,0} )/\sqrt{1+\xi^2}$ and for some value $\xi \simeq 0.01$ (found numerically to provide the best robustness%
\footnote{Somewhat surprisingly, taking $\xi = 0$ would generate a causally separable D-POVM. One can thus see that the preparation of the quantum input states needs to be quite faithful; while being relatively robust to the noise in the process matrix $W_\text{FAB}$, our certification here is not very robust to noise in the quantum inputs.}%
). 

The induced D-POVM $\mathbb{E}^{\tilde{A}\tilde{B}}(r) = \big(E_{a,b}^{\tilde{A}\tilde{B}}(r) = (M_a^{\tilde{A}A} \otimes M_b^{\tilde{B}B}) * W_\text{FAB}(r)\big)_{a,b}$ is then easily obtained, and using the characterisations of Sec.~\ref{app_cones_N2} we could find (numerically, using the SDP solver Mosek~\cite{mosek} with CVX~\cite{cvx} or YALMIP~\cite{yalmip}) that it is causally nonseparable for $r\lesssim 0.113$, as claimed above.  It remains an open question, whether our choice of instruments above is optimal; it could well be, in particular, that increasing the dimension of the quantum inputs could improve the robustness to noise, for certifying the causal nonseparability of $W_\text{FAB}$ in a SDI-QI manner.

\medskip

Recall from Ref.~\cite{feix16} that despite being causally nonseparable, $W_\text{FAB}$ cannot by itself violate a causal inequality: it is in this sense ``causal''. The argument to show that it can only generate causal correlations is based on the fact that it generates (when considering all possible instruments) the same set of correlations $p(a,b|x,y) = (M_{a|x}^A\otimes M_{b|y}^B) * W_\text{FAB} = (M_{a|x}^A\otimes (M_{b|y}^B)^T) * W_\text{FAB}^{T_B}$ as $W_\text{FAB}^{T_B}$ (where ${}^{T_B}$ denotes the partial transpose over Bob's spaces), which is causally separable. This argument can however no longer be applied to the induced D-POVMs: one may still write $E_{a,b}^{\tilde{A}\tilde{B}} = (M_a^{\tilde{A}A}\otimes M_b^{\tilde{B}B}) * W_\text{FAB} = (M_a^{\tilde{A}A}\otimes (M_b^{\tilde{B}B})^{T_B}) * W_\text{FAB}^{T_B}$, but here the $(M_b^{\tilde{B}B})^{T_B}$'s may not be PSD and may thus not be valid instrument elements---as it is indeed the case for our choice in Eq.~\eqref{eq:instr_W_FAB} above.

The partial transpose argument can also no longer be applied when some entangled ancillary state is attached to $W_\text{FAB}$. And indeed, it was found in Ref.~\cite{feix16} that $W_\text{FAB}$ is not ``extensibly causal'': $W_\text{FAB}$ does allow for some (small) causal inequality violation when extended by some ancillary state.
Hence, the approach and instruments of Sec.~\ref{app:noncausal_imply_nonsepDPOVM} could also be used to generate a causally nonseparable D-POVM from $W_\text{FAB}$. This non-extensible-causality was however not found to be very robust to noise: such ``activation'' of noncausality by entanglement was only found for noise values much smaller than the robustness $r\simeq 0.113$ found above,%
\footnote{More specifically, Ref.~\cite{feix16} considered the noisy ``extended'' process $(1{-}\kappa)\,W_\text{FAB} \otimes \rho^{A'B'} + \kappa \,\id^{ABA'B'}/64$, for some maximally entangled 2-ququart ancillary state $\rho^{A'B'}$, and only found causal inequality violations with this process matrix for $\kappa\lesssim 3.3\times 10^{-4}$. From this, one may expect $W_\text{FAB}(r)$ to also exhibit non-extensible-causality for $r$ of the order of $10^{-4}$ -- $10^{-3}$ only.}
so that our approach above to certify the causal nonseparability of $W_\text{FAB}$, with our choice of instruments in Eq.~\eqref{eq:instr_W_FAB}, is much more robust to noise than via its non-extensible-causality.

\subsection{SDI-QI and MDCI causal nonseparability of the quantum switch}
\label{app:QSwitch}

The quantum switch~\cite{chiribella13}, with a qubit ``target system'' initialised in the state $\ket{0}$ and traced out at the end, and a qubit ``control system'' initialised in the state $\ket{+} = \frac{1}{\sqrt{2}}(\ket{0}+\ket{1})$, can be written as a (2+$F$)-partite process matrix as follows~\cite{araujo15,oreshkov16}:
\begin{align}
W_\text{QS} = & \Tr_{F_\text{t}} \ketbra{w_\text{QS}}{w_\text{QS}}^{ABF_\text{t}F} \qquad \text{with} \notag \\[2mm]
\ket{w_\text{QS}}^{ABF_\text{t}F} = & {\textstyle \frac{1}{\sqrt{2}}} \ket{0}^{A_I}\dket{\id}^{A_OB_I}\dket{\id}^{B_OF_\text{t}}\ket{0}^F \notag \\
& + {\textstyle \frac{1}{\sqrt{2}}} \ket{0}^{B_I}\dket{\id}^{B_OA_I}\dket{\id}^{A_OF_\text{t}}\ket{1}^F
\end{align}
(and with implicit tensor products).

We consider in this paper a ``depolarised'' version of the quantum switch, obtained by mixing it with some fully white noise described by the process matrix $\id^\circ = \id^{ABF}/8$:
\begin{align}
W_\text{QS}(r) = {\textstyle \frac{1}{1+r}}(W_\text{QS} + r \, \id^\circ), \label{eq:WQSr_app}
\end{align}
for some noise parameter $r\ge 0$, as in Eq.~\eqref{eq:WQSr} of the main text.

\subsubsection{SDI-QI certification}

For the choice of instruments given in Eq.~\eqref{eq:Ms_QS} (with 2-dimensional quantum input states for Alice and Bob, and no quantum input for Fiona), we obtain the D-POVM
\begin{align}
\mathbb{E}_\text{QS}(r) = {\textstyle \frac{1}{1+r}}(\mathbb{E}_\text{QS} + r \, \mathbb{E}^\circ)
\end{align}
with
\begin{align}
\mathbb{E}_\text{QS} & = \big(E_{a,b,f}^{\tilde{A}\tilde{B}} = \Tr_{F_\text{t}} \ketbra{e_{a,b,f}}{e_{a,b,f}}\big)_{a,b=0,1;f=\pm} , \notag \\
\ket{e_{a,b,\pm}} & = \frac12 \big( \delta_{a,0}\ket{b}^{\tilde{A}} \dket{\id}^{\tilde{B}F_\text{t}} \pm \delta_{b,0}\ket{a}^{\tilde{B}} \dket{\id}^{\tilde{A}F_\text{t}} \big), \notag \\[1mm]
\mathbb{E}^\circ & = \big(E_{a,b,f}^{\circ\, \tilde{A}\tilde{B}} = \frac18 \id^{\tilde{A}\tilde{B}} \big)_{a,b,f}.
\end{align}
We show here that $\mathbb{E}_\text{QS}(r)$ above is causally nonseparable for $r < 2- 2\sqrt{\frac{2}{3}} \simeq 0.367$, by providing an explicit witness of causal nonseparability.

\medskip

Let us define for that $\mathbb{S}_\text{QS} = (S_{a,b,f}^{\tilde{A}\tilde{B}} = S_{\text{PSD};a,b,f}^{\tilde{A}\prec\tilde{B}\prec\tilde{F}} + S_{a,b}^{\tilde{A}\prec\tilde{B}\prec\tilde{F}} + S_a^{\tilde{A}\prec\tilde{B}\prec\tilde{F}} + S^{\tilde{A}\prec\tilde{B}\prec\tilde{F}} = S_{\text{PSD};a,b,f}^{\tilde{B}\prec\tilde{A}\prec\tilde{F}} + S_{a,b}^{\tilde{B}\prec\tilde{A}\prec\tilde{F}} + S_b^{\tilde{B}\prec\tilde{A}\prec\tilde{F}} + S^{\tilde{B}\prec\tilde{A}\prec\tilde{F}})_{a,b=0,1;f=\pm}$ with
\begin{align}
S_{\text{PSD};0,0,\pm}^{\tilde{A}\prec\tilde{B}\prec\tilde{F}} & = \ketbra{\psi_{v,u}^\mp}{\psi_{v,u}^\mp}, \quad S_{\text{PSD};0,1,f}^{\tilde{A}\prec\tilde{B}\prec\tilde{F}} = (u{-}v) \ketbra{00}{00}, \notag \\
S_{\text{PSD};1,0,f}^{\tilde{A}\prec\tilde{B}\prec\tilde{F}} & = (u{-}v) \ketbra{10}{10}, \quad S_{\text{PSD};1,1,f}^{\tilde{A}\prec\tilde{B}\prec\tilde{F}} = 0, \notag \\
S_0^{\tilde{A}\prec\tilde{B}\prec\tilde{F}} & = (u{-}v) ( \ketbra{01}{01} - \ketbra{00}{00}), \notag \\
S_{a,b}^{\tilde{A}\prec\tilde{B}\prec\tilde{F}} & = S_1^{\tilde{A}\prec\tilde{B}\prec\tilde{F}} = S^{\tilde{A}\prec\tilde{B}\prec\tilde{F}} = 0, \notag \\[1mm]
S_{\text{PSD};0,0,\pm}^{\tilde{B}\prec\tilde{A}\prec\tilde{F}} & = \ketbra{\psi_{u,v}^\mp}{\psi_{u,v}^\mp}, \quad S_{\text{PSD};1,0,f}^{\tilde{B}\prec\tilde{A}\prec\tilde{F}} = (u{-}v) \ketbra{00}{00}, \notag \\
S_{\text{PSD};0,1,f}^{\tilde{B}\prec\tilde{A}\prec\tilde{F}} & = (u{-}v) \ketbra{01}{01}, \quad S_{\text{PSD};1,1,f}^{\tilde{B}\prec\tilde{A}\prec\tilde{F}} = 0, \notag \\
S_0^{\tilde{B}\prec\tilde{A}\prec\tilde{F}} & = (u{-}v) ( \ketbra{10}{10} - \ketbra{00}{00}), \notag \\
S_{a,b}^{\tilde{B}\prec\tilde{A}\prec\tilde{F}} & = S_1^{\tilde{B}\prec\tilde{A}\prec\tilde{F}} = S^{\tilde{B}\prec\tilde{A}\prec\tilde{F}} = 0,
\end{align}
with $u = \frac{\sqrt{6}+2}{3}$, $v = \sqrt{6}-2$, $\ket{\psi_{u,v}^\pm} = \sqrt{u} \ket{01} \pm \sqrt{v} \ket{10}$, $\ket{\psi_{v,u}^\pm} = \sqrt{v} \ket{01} \pm \sqrt{u} \ket{10}$, and with all $S$ matrices above being written in the space $\L(\HS^{\tilde{A}\tilde{B}})$ (since $\HS^{\tilde{F}}$ is trivial here).

Recall that in the (2+$F$)-partite scenario under consideration here, the set of causal witnesses for D-POVMs are described by Eqs.~\eqref{eq:dual_cone_2F} and~\eqref{eq:dual_cone_2F_2}. It is easily verified that the matrices above satisfy the required constraints, so that $\mathbb{S}_\text{QS}$ defines a valid witness.

We find $\mathbb{S}_\text{QS} * \mathbb{E}_\text{QS}(r) = \frac{1}{1+r}(\mathbb{S}_\text{QS} * \mathbb{E}_\text{QS} + r \, \mathbb{S}_\text{QS} * \mathbb{E}^\circ)$ with $\mathbb{S}_\text{QS} * \mathbb{E}_\text{QS} = -( \sqrt{uv}-v) = -\Big( 2- 2\sqrt{\frac{2}{3}} \Big)$ and $\mathbb{S}_\text{QS} * \mathbb{E}^\circ = 1$ (as in Eq.~\eqref{eq:dual}), so that $\mathbb{S}_\text{QS} * \mathbb{E}_\text{QS}(r) < 0$---which indeed implies that $\mathbb{E}_\text{QS}(r)$ is causally nonseparable, and that the noisy quantum switch can thus be certified in a SDI-QI manner (by reconstructing the witness from some correlations $P(a,b|\rho_x^{\tilde{A}},\rho_y^{\tilde{B}})$, as discussed in the main text)---for $r < 2- 2\sqrt{\frac{2}{3}} \simeq 0.367$.

\medskip

The choice of instruments in Eq.~\eqref{eq:Ms_QS}, as considered here, is the best one we found, that gave the largest noise robustness; the witness above was initially obtained numerically, using CVX~\cite{cvx} and Mosek~\cite{mosek}.
Recall that the depolarised quantum switch $W_\text{QS}(r)$ is causally nonseparable for $r \lesssim 1.576$~\cite{branciard16a}: there thus remains an important gap (for $0.367 \lesssim r \lesssim 1.576$), for which we do not know if the quantum switch can be certified to be causally nonseparable in a SDI-QI manner. It could be that some better choice of instruments would allow one to partially close this gap---although from our various tests, we conjecture that some nonzero gap would still remain, i.e., that for a certain range of values $r$, $W_\text{QS}(r)$ is causally nonseparable but that this cannot be certified in a SDI-QI manner.

\subsubsection{MDCI certification}

It was shown in Ref.~\cite{bavaresco19} that the depolarised quantum switch~\eqref{eq:WQSr_app} is ``TTU-noncausal'' (see Sec.~\ref{app:anyTTU}) for $r \lesssim 1.319$ (or in terms of the noise parameter used in~\cite{bavaresco19}, for $\eta = \frac{r}{1+r} \lesssim 0.5687$). Indeed for such values of $r$, the TTU-assemblage $(w_{f=\pm}^{AB} = \ketbra{\pm}{\pm}^F * W_\text{QS}(r))_{f=\pm}$ induced when Fiona measures her qubit in the (fixed) $\{\ket{\pm}\}$ basis was found to be noncausal.

Accordingly, as discussed in Sec.~\ref{app:MDCI_2F_TTU}, this implies that the noisy quantum switch can be certified to be causally nonseparable in a MDCI-MDCI-DI scenario (i.e., with trusted quantum inputs for Alice and Bob and with their instruments of the form of Eq.~\eqref{eq:cstr_instr_v0}, but with a fully untrusted device for Fiona) for all $r \lesssim 1.319$. 
Hence, imposing the structure of Eq.~\eqref{eq:cstr_instr_v0} to Alice and Bob's instruments, allows one to partially close the gap mentioned above.

\medskip

It was also shown in Ref.~\cite{bavaresco19} that $W_\text{QS}(r)$ is ``TUU-noncausal'' (see Sec.~\ref{app:anyTUU}) for $r \lesssim 0.194$ (or $\eta = \frac{r}{1+r} \lesssim 0.1621$). To obtain a noncausal TUU-assemblage $(w_{b,f=\pm|y}^{AB} = (M_{b|y}^B\otimes\ketbra{\pm}{\pm}^F) * W_\text{QS}(r))_{b,f=\pm,y}$ for such values of $r$, Bavaresco \emph{et al.} considered two different instruments for Bob (labeled by $y=0,1$), given by $(M_{b=0,1|y=0}^B = \ketbra{b}{b}^{B_I} \otimes \ketbra{b}{b}^{B_O})_{b=0,1}$ and $(M_{b=\pm|y=1}^B = \ketbra{\pm}{\pm}^{B_I} \otimes \ketbra{\pm}{\pm}^{B_O})_{b=\pm}$, with Fiona still measuring in the fixed $\{\ket{\pm}\}$ basis.

Accordingly, as discussed in Sec.~\ref{app:MDCI_2F_TUU}, this implies that the noisy quantum switch can be certified to be causally nonseparable in a MDCI-DI-DI scenario (i.e., with trusted quantum inputs for Alice only, and her instruments of the form of Eq.~\eqref{eq:cstr_instr_v0}, but with fully untrusted devices for Bob and Fiona) for all $r \lesssim 0.194$.

\medskip

In both cases above, it remains an open question, whether the choice of instruments for Fiona and Bob (in the second case) is optimal, or whether different choices may allow one to certify the causal nonseparability of the quantum switch for some larger range of the noise parameter $r$.

\end{document}